\documentclass[aps,preprintnumbers, prd, twocolumn, floatfix, preprintnumbers, letterpaper, amsmath, amssymb, superscriptaddress,nofootinbib]{revtex4}
\usepackage{amssymb, amsmath, bm, dcolumn, epsf, graphicx, latexsym, mathbbol, slashed, simplewick}
\usepackage{graphicx}
\usepackage{subfigure}
\usepackage{dcolumn}
\usepackage{bm}
\usepackage{amssymb}
\usepackage{color}
\usepackage{multirow}
\usepackage{makecell}
\usepackage[colorlinks,linkcolor=blue,citecolor=blue,urlcolor=blue]{hyperref}
\usepackage{upgreek}
\usepackage{accents}
\usepackage{verbatim}
\usepackage{mathrsfs}
\usepackage{amsmath}
\def\f{\frac}
\def\M{\mathcal M}
\def\E{\mathcal E}

 \def\tskip{\setlength{\tskip}{5pt}}
\def\colwidth{\setlength{\colwidth}{3.5in}}

\def\prd{Phys. Rev. D}

\def\prl{Phys. Rev. Lett.~}

\def\apj{Astrophys. J.~}
\def\apjl{Astrophys. J. Lett.~}

\def\mnras{Mon. Not. Roy. Astron. Soc.~}

\newcommand{\be}{\begin{equation}}
\newcommand{\ee}{\end{equation}}

\newcommand{\ep}{\epsilon}

\newcommand{\om}{\omega}
\newcommand{\ph}{\phi_{\rm VEV}}
\newcommand{\ac}{\accentset}
\def\Mpl{M_{\mathrm{Pl}}}
\newcommand{\A}{A_{\rm VEV}}
\newcommand{\V}{V_{\rm VEV}}

\begin{document}
%\title{Constraints on screened modified gravity from binary pulsar observations}
%\title{Constraints on screened modified gravity from gravitational waves}
\title{Gravitational radiation from compact binary systems in screened modified gravity}
%\title{Scalar radiation from compact binary systems in screened modified gravity}
%\title{Pulsar constraints on screened modified gravity}
%\title{Gravitational radiation and Solar System constraints on screened modified gravity}

\author{Xing Zhang}
\email[]{starzhx@mail.ustc.edu.cn}
\author{Tan Liu}
\email[]{lewton@mail.ustc.edu.cn}
\author{Wen Zhao}
\email[]{wzhao7@ustc.edu.cn}
\affiliation{CAS Key Laboratory for Researches in Galaxies and Cosmology, Department of Astronomy, University of Science and Technology of China, Chinese Academy of Sciences, Hefei, Anhui 230026, China}

\date{\today}

\begin{abstract}
Screened modified gravity (SMG) is a kind of scalar-tensor theory with screening mechanisms, which can suppress the fifth force in dense regions and allow theories to evade the solar system and laboratory tests. In this paper, we investigate how the screening mechanisms in SMG affect the gravitational radiation damping effects, calculate in detail the rate of the energy loss due to the emission of tensor and scalar gravitational radiations, and derive their contributions to the change in the orbital period of the binary system. We find that the scalar radiation depends on the screened parameters and the propagation speed of scalar waves, and the scalar dipole radiation dominates the orbital decay of the binary system. For strongly self-gravitating bodies, all effects of scalar sector are strongly suppressed by the screening mechanisms in SMG. By comparing our results to observations of binary system PSR J1738+0333, we place the stringent constraints on the screening mechanisms in SMG. As an application of these results, we focus on three specific models of SMG (chameleon, symmetron, and dilaton), and derive the constraints on the model parameters, respectively.
\end{abstract}


\maketitle

\section{Introduction} \label{section1}

Einstein's theory of General Relativity (GR) has been very successful at interpreting gravity on a huge range of scales, from submillimeter scale tests in the laboratory \cite{Hoyle2001p1418,Adelberger2001}, to solar system \cite{Will1993p,Will2014p4} and binary pulsar \cite{Stairs2003p5,Wex2014,Manchester2015p1530018,Kramer2016p1630029} tests. Nevertheless, GR is known to be incomplete in the ultraviolet regime where it should be replaced by a (still unknown) quantum theory of gravity \cite{Kiefer2007p}.  Also, within the framework of GR, in order to explain the observations on the infrared cosmological scales, the dark ingredients (dark matter and dark energy) \cite{Cline2013p1} were introduced as the supplementary material in our Universe. Therefore, alternative theory of gravity is the direction that is supposed to be worth a try. In addition, the majority of tests of GR only verify the effects of the conservative sector of GR in the weak-field and low energy regimes \cite{Will1993p,Will2014p4}. Gravitational waves (GWs) provide the excellent opportunity to perform quantitative tests of dissipative sector and strong-field dynamics of gravity theories. The first indirect detection of GWs is based on the observations of orbital decay of binary pulsar system \cite{Taylor1994p711}. In September 14, 2015, the first direct GW signal GW150914 was observed by LIGO, which was produced by the coalescence of two stellar-mass black holes \cite{Abbott2016p61102}. In order to better understand gravity and fundamental physics from these observations, it is important to clarify the corresponding predictions from GR and alternative theories of gravity \cite{Yunes2013p9,Clifton2012p1}. For these reasons, the study of gravitational radiation in alternative theories of gravity has become an important issue.

Another motivation for the research on gravity theories is the following argument: Scientists can never truly ``prove" that a theory (e.g. GR) is correct, but rather all we do is disprove, or more accurately constrain, alternative hypothesis. The theory that remain and cannot be disproven by observations becomes the {\emph{status quo}} \cite{karl}. Indeed, this is the case today for Einstein's theory of GR. So, even for the verification of GR theory, we should also investigate the prediction of alternative theories, and compare them with the prediction of GR.  Actually, this has been subjected to a battery of tests through solar system \cite{Will2014p4,Ni2016p1630003}, binary pulsar \cite{Stairs2003p5,Wex2014,Manchester2015p1530018,Kramer2016p1630029}, gravitational waves in the binary black holes \cite{Yunes2016} and cosmological observations \cite{cosmologytests}.

A natural alternative to GR is scalar-tensor theory \cite{Fujii2003p,Faraoni2004p,Damour1992p2093}, which invokes a conformal coupling between matter and an underlying scalar field, besides the standard space-time metric tensor. Scalar-tensor theory can not only be shown to be equivalent to several phenomenological gravity theories (e.g. $f(R)$ gravity \cite{Sotiriou2010p451,DeFelice2010p3}), but also be justified by the low energy limit of string theory or supergravity \cite{Becker2006p, Damour1994p532, Damour1994p1171, Damour2002p46007}. Moreover, scalar fields are also widely used in modern cosmology (e.g. quintessence \cite{Caldwell1998p1582} and inflation \cite{Guth1981p347}). The coupling between scalar field and matter leads to the scalar force (fifth force), and current experimental constraints \cite{Adelberger2009p102,Williams2012p184004} require that the fifth force must be screened in high density environments. Presently, there are three main screening mechanisms in scalar-tensor gravity: chameleon \cite{Khoury2004p44026,Khoury2004p171104,Gubser2004p104001}, symmetron \cite{Hinterbichler2010p231301,Hinterbichler2011p103521,Davis2012p61}, and dilaton \cite{Damour1994p532,Damour1994p1171,Brax2010p63519}\footnote{The screen mechanism can also be realized by the non-linearities in the kinetic term $p(\phi,X)$ of scalar field \cite{Vainshtein}, which is not considered in the present article. }. These three mechanisms can be described within a unified theoretical framework called screened modified gravity (SMG) \cite{Brax2012p44015}. SMG is a class of scalar-tensor theory with screening mechanisms, which is described by a bare potential $V(\phi)$ and a conformal coupling function $A(\phi)$ in scalar-tensor theory. The motion of scalar field is governed by an effective potential defined through $V(\phi)$ and $A(\phi)$. In order that SMG can generate a screening effect to suppress the fifth force in high density environments, the effective potential must have a minimum \cite{Brax2012p44015}, which can be naturally understood as a physical vacuum. Around this physical vacuum, the scalar field acquires an effective mass, which increases as the ambient density increases. Therefore, the scalar field can be screened in high density regions (small scales), where the range of the fifth force (scalar force) is so short that it cannot be detected within current experimental accuracy \cite{Khoury2004p171104,Gubser2004p104001}. Whereas in low density regions (galactic and cosmological scales), the long-range fifth force may affect galactic dynamics \cite{Gronke2015p123,Schmidt2010p103002}, and the scalar potential can play the role of dark energy to accelerate the expansion of the Universe \cite{Khoury2004p44026,Hinterbichler2011p103521}.

The salient feature of SMG is the screening mechanism, which can suppress the fifth force and allow theories to evade the solar system tests. In previous work \cite{Zhang2016p124003}, we have investigated the screening mechanisms for the SMG with a general potential $V(\phi)$ and coupling function $A(\phi)$, and calculated the parametrized post-Newtonian (PPN) parameters, the effective gravitational constant, and the effective cosmological constant. Based on these, we derived the constraints on the model parameters by combining the observations on solar system and cosmological scales. As an extension of this issue, in this paper we investigate how the screening mechanisms in SMG affect the gravitational radiation damping of compact binary systems. We calculate in detail the rate of the energy loss due to the emission of tensor and scalar gravitational radiations (including monopole, dipole, quadrupole, and dipole-octupole radiations) from compact binary systems in SMG. We pay particular attention on dipole radiation, which is generally stronger than GR's quadrupole radiation, and might dominate the orbital decay of the binary system.

In earlier work, Eardley \cite{Eardley1975p59} was the first to point out the existence of dipole gravitational radiation from self-gravitating bodies in the Brans-Dicke gravity, and Will {\it et al.} \cite{will1989} and Alsing {\it et al.} \cite{Alsing2012p64041} placed the pulsar constraints on the massless and massive Brans-Dicke gravity, respectively. Damour and Esposito-Far\`{e}se \cite{Damour1992p2093} derived the tensor and scalar gravitational radiation fluxes in the massless multi-scalar-tensor theories. However, these theories do not have screening effects. Brax {\it et al.} \cite{Brax2014p225001} investigated how the cosmological evolution of the scalar field in SMG results in the emission of scalar radiation. However, he did not consider that the objects spiral into each other results in the emission of gravitational radiation, as a complement, in this paper we focus on this case.

In general, in any theory of gravity (including GR), GWs emission depends not only on the dissipative sector of the theory which regulates how fast the binary system loses energy, but also on the conservative sector of the theory which regulates the orbital dynamics of the system. In alternative theories of gravity, in general, both the conservative and dissipative sectors are modified relative to GR. In order to understand the effects of the dissipative sector of the theory, we first need to consider the modifications to the conservative sector.

In the conservative sector of SMG, we study the impact of the screening mechanism on the orbital dynamics of compact binary systems, which can be effectively described by the point-particle action with $\phi$-dependent mass introduced by Eardley \cite{Eardley1975p59}. In alternative theories (including SMG), the orbital dynamics is generally modified by the additional fields controlled by the sensitivities \cite{Eardley1975p59}, which characterize how the gravitational binding energy of the object responds to its motion relative to the additional fields. In the weak-field limit around the Minkowski background and the scalar background (the vacuum expectation value (VEV) of the scalar field), making use of the post-Newtonian (PN) formalism, we solve the PN equations for the massless tensor and massive scalar fields in the near zone. By comparing this scalar field solution with our previous result \cite{Zhang2016p124003} obtained by the method of matching the internal and external solutions, we find that the first sensitivity of the object is completely equivalent to its screened parameter. We utilize these PN solutions to derive the equations of motion for compact binary systems by adopting the method of Einstein, Infeld and Hoffmann (EIH) \cite{Einstein1938p65}. It turns out that the equations of motion at Newtonian order violate not only the weak equivalence principle (WEP) but also the gravitational inverse-square law. However, in the near zone the inverse-square law approximately holds, which guarantees the Kepler's third law in this scale.

In the dissipative sector of SMG, we solve the wave equations for the massless tensor and massive scalar fields in the wave zone, and derive the energy fluxes carried by the tensor and scalar modes by investigating the conserved charges and currents in this theory. We find that the tensor and scalar modes carry away energy from the source starting at quadrupole and monopole orders, respectively. These emerge as the consequences of the facts that the tensor and scalar gravitons are respectively the massless spin-2 and massive spin-0 particles \cite{Maggiore2007p}. In the tensor sector of SMG, like in GR, the tensor gravitational charge is the mass itself. Consequently, the conservations of mass and momentum forbid monopole and dipole tensor radiations, and the tensor quadrupole radiation in SMG behaves similar to that in GR at leading PN order. In the scalar sector of SMG, the scalar radiation (including monopole, dipole, quadrupole, and dipole-octupole radiations) depends strongly on the screened parameter, which acts as the scalar gravitational charge in the theory. There is no scalar monopole radiation contribution to leading order in the quasi-circular orbit case. The (scalar) dipole radiation is present in SMG or in other alternative theories of gravity. This is because that the violation of the WEP in these theories leads to the difference between the two centers of gravitational and inertial masses of the system, which induces a time-varying dipole moment that emits radiation as the objects spiral into each other. The dipole-octupole cross term appearing in the scalar radiation is the negative modification to the energy flux at the same PN order as the quadrupole radiation contribution. In alternative theories (including SMG), the dipole radiation generally depends on the difference in sensitivities (screened parameters in SMG), since the conservation of momentum turns the `charge' dipole moment into the form of the difference in sensitivities. In SMG, the scalar radiation also depends on the propagation speed of the massive scalar particle, which satisfies the relativistic dispersion relation. This result shows that in SMG the scalar GWs can be emitted (i.e., scalar mode is excited) if and only if the frequency (energy) of scalar mode is greater than its mass.

In this paper, we pay particular attention on dipole radiation, which is generally stronger than GR's quadrupole radiation and leads to a strong modification on the orbital evolution of compact binary systems. However, in SMG, we find that the scalar dipole radiation, as well as the other modifications in the conservative and dissipative sectors, are all suppressed by the screening mechanisms, and thus the deviations from GR become small for strongly gravitating bodies (such as white dwarfs and neutron stars). Since in SMG, the screened parameter (or sensitivity) of the object is inversely proportional to its surface gravitational potential, which induces that the SMG is completely different from other alternative theories without screening mechanisms \cite{Damour1993p2220,Damour1996p1474a}, and possibly passes the accurate tests in binary systems \cite{Antoniadis2013p6131,Freire2012p3328}. Finally, we obtain the stringent bounds on the screened parameter (and scalar field VEV) by comparing our results for the orbital period decay rate to the observations of quasi-circular binary system PSR J1738+0333 \cite{Freire2012p3328}.  As an application of these results, we focus on three specific models of SMG (chameleon, symmetron, and dilaton), and derive the constraints on the model parameters, respectively.

This paper is organized as follows. In Sec. \ref{section2}, we display the action for SMG and derive the field equations and their weak-field limit. In Sec. \ref{section3}, we focus on the conservative sector of SMG, solve the PN equations for the tensor and scalar fields, and investigate the orbital dynamics of binary systems. In Sec. \ref{section4}, we focus on the dissipative sector of SMG, calculate the rate of the energy loss due to the tensor and scalar gravitational radiations, and derive their contributions to the change of the orbital period. In Sec. \ref{section5}, we apply our results to three specific models of SMG (chameleon, symmetron, and dilaton), and derive the constrains on these models by the current observations. We conclude in Sec. \ref{section6} with a summary and discussion. %% In Appendix \ref{appendix_a}, we review the derivation of the scalar exterior solution. Finally, Appendix \ref{appendix_b} outlines a step-by-step derivation of the PN expansion of the metric tensor.

Throughout this paper, the metric convention is chosen as $(-,+,+,+)$, and Greek indices ($\mu,\nu,\cdots$) run over $0,1,2,3$. We set the units to $c=\hbar=1$, and therefore the reduced Planck mass is $M_\text{Pl} = \sqrt{1/8 \pi G}$, where $G$ is the gravitational constant.

\section{Screened Modified Gravity}\label{section2}

\subsection{The action}
Screened modified gravity (SMG) is a class of scalar-tensor theory with screening mechanisms, which can suppress the fifth force in dense regions and pass the solar system tests \cite{Zhang2016p124003}. A general scalar-tensor gravity with two arbitrary functions is given by the following action in the Einstein frame \cite{Damour1992p2093,Brax2012p44015}:
\begin{align} \label{action_0}
\begin{split}
S=&\int d^4x\sqrt{-g}\left[\frac{\Mpl^2}{2}R-\frac12(\nabla\phi)^2-V(\phi)\right]
\\&+S_m\left[A^2(\phi) g_{\mu\nu},\,\psi_m^{(i)}\right],
\end{split}
\end{align}
where $g$ is the determinant of the Einstein frame metric $g_{\mu\nu}$, $R$ is the Ricci scalar, $\psi_m^{(i)}$ are various matter fields labeled by $i$\,. The bare potential $V(\phi)$ characterizes the scalar self-interaction, which has three main effects in the theory: First, it can play the role of dark energy to accelerate the expansion of the universe at late times. Second, it endows the scalar field with mass. Finally, it may introduce nonlinearities into the scalar dynamics. $A(\phi)$ is a conformal coupling function characterizing the interaction between the scalar and matter fields, which induces the fifth force (scalar force) in the theory. In the Einstein frame, the scalar field interacts directly with the matter field through the conformal coupling function $A(\phi)$. In the Jordan frame, the matter field couples to the Jordan frame metric $\widetilde{g}_{\mu\nu}$ through a conformal rescaling of the Einstein frame metric $g_{\mu\nu}$ as $\widetilde{g}_{\mu\nu}=A^2(\phi)g_{\mu\nu}$ \cite{Faraoni2007p23501,Postma2014p103516a}. The coupling function $A(\phi)$ is usually different for different matter fields $\psi_m^{(i)}$, but for simplicity we assume that all matter fields couple in the same way to the scalar field with a universal coupling function $A(\phi)$.

In general, the scalar field equation is Klein-Gordon equation $\square_g\phi={\partial V_{\rm eff}}/{\partial \phi}$ in Eq. \eqref{scalar_eom}. The scalar field is governed by the effective potential $V_{\text {eff}}(\phi)$ defined in \eqref{Veff2}, which depends on the bare potential $V(\phi)$ and coupling function $A(\phi)$. The shape of the effective potential determines the behavior of the scalar field. For suitably chosen functions $V(\phi)$ and $A(\phi)$, the effective potential $V_{\rm eff}(\phi)$ can have a minimum, i.e., the scalar field has a physical vacuum \cite{Brax2012p44015,Zhang2016p124003},
\begin{align}\label{mass_eff}
\frac{\rm d V_{\rm eff}}{\rm d\phi}\bigg|_{\phi_{\rm min}}=0\,,\quad~    m^2_{\rm eff}\equiv \frac{\rm d^2 V_{\rm eff}}{\rm d\phi^2}\bigg|_{\phi_{\rm min}}>0\,.
\end{align}
Around this minimum (physical vacuum), the scalar field acquires an effective mass which increases as the ambient density increases. Therefore, the scalar field can be screened inside matter overdensities (high density), where the fifth force range is so short that it cannot be detected within current experimental accuracy. This kind of scalar-tensor gravity with screening mechanism is often called screened modified gravity \cite{Brax2012p44015,Brax2014p23505,Brax2014p225001}, which can generate the screening effect to suppress the fifth force in high density environments and pass the solar system and laboratory tests. There are many SMG models in the market, including the chameleon, symmetron and dilaton models \cite{Brax2012p44015}, in which the functions $V(\phi)$ and $A(\phi)$ are chosen as the specific forms.

\subsection{Point-particle action of compact \\objects and field equations}
GR satisfies exactly the strong equivalence principle (SEP) which leads to a happy property called the ``effacement'' principle \cite{Damour1987p128}. This principle states that the internal structure of strongly self-gravitating bodies is ``effaced'' and their dynamics and radiation depend only on their masses and spins (for simplicity we do not consider the spin effects in this article). However, the effacement principle does not hold in alternative theories of gravity like scalar-tensor gravity. In scalar-tensor theory, the inertial mass and internal structure of a strongly self-gravitating body depend on the local scalar field (i.e., the local gravitational coupling ``constant''), which may act back on the motion of the body and lead to violation of the SEP. In general, so long as the compact objects are far enough from each other, their motion can be effectively described through point particles with the composition dependent effects encapsulated in nonstandard couplings in the particle action. Eardley \cite{Eardley1975p59} first showed that these effects could be accounted for by supposing the mass of the body as a function of the scalar field, such that the matter action for a system of point-like masses can be written as
\begin{align}
\label{matter_action}
\begin{split}
S_m=&-\sum_a\int m_a(\phi)d\tau_a\,,
\end{split}
\end{align}
where $m_a(\phi)$ is the $\phi$-dependent mass of the $a$-th point-particle, and $\tau_a$ is its proper time measured along its worldline $x^\lambda_a$. From this action we can clearly observe that the WEP is violated, since the scalar field depends on position, the mass becomes position-dependent, and the variation of $S_m$ does not yield the geodesic equation. Using the definition of $T^{\mu\nu} \equiv(2/\sqrt{-g})\delta S_m/\delta g_{\mu\nu}$, the energy-momentum tensor of matter $T^{\mu\nu}$ and its trace $T$ hence take the form
\begin{align}\label{Tuv_matter}
T^{\mu\nu}(x,\phi)=(-g)^{-1/2}\sum_{a}m_a(\phi)\frac{u^{\mu}_{a}u^{\nu}_{a}}{u^{0}_{a}}\delta^3(\mathbf{r}-\mathbf{r}_a(t))\,,
\end{align}
\begin{align}\label{Tm_trace}
T(x,\phi)=-(-g)^{-1/2}\sum_{a}\frac{m_a(\phi)}{u^{0}_{a}}\delta^3(\mathbf{r}-\mathbf{r}_a(t))\,,
\end{align}
where $u^{\mu}_a$ is four-velocity of the $a$-th point-particle, and $\delta^3$ is the three-dimensional Dirac delta function.

The full action for a system of compact objects is now given by
\begin{align} \label{action_1}
\begin{split}
S=&\int d^4x\sqrt{-g}\left[\frac{\Mpl^2}{2}R-\frac12(\nabla\phi)^2-V(\phi)\right]
\\&-\sum_a\int m_a(\phi)d\tau_a\,.
\end{split}
\end{align}
The variation of the action \eqref{action_1} with respect to the tensor field and the scalar field yields the tensor field equation of motion (EOM)
%\begin{subequations}
\begin{align}
G_{\mu\nu}\label{tensor_eom}
= 8 \pi G \left[T_{\mu\nu}(x,\phi)+T_{\phi\mu\nu}(\phi)\right]\,,
\end{align}
and the scalar field EOM
\begin{align}
\square_g\phi\label{scalar_eom}
=\frac{\partial V_{\rm eff}(\phi)}{{\partial }\phi}\,,
\end{align}
%\end{subequations}
where $\square_g\equiv(-g)^{-1/2}\partial_{\nu}\left((-g)^{1/2}g^{\mu\nu}\partial_{\mu}\right)$ is the curved space d'Alembertian. Note that, $G$ is the bare gravitational constant, and it is related to the Newtonian gravitational constant measured with Cavendish-type experiments through Eq.\eqref{G_12}\,.
Here, $G_{\mu\nu}$ is the Einstein tensor, $T_{\mu\nu}(x,\phi)$ is the matter energy-momentum tensor given in Eq. \eqref{Tuv_matter},
\begin{align}\label{Tuv_phi}
T_{\phi\mu\nu}(\phi)=\partial_\mu\phi\partial_\nu\phi-g_{\mu\nu}\left[\frac{1}{2}(\partial\phi)^2+V(\phi)\right]
\end{align}
is the scalar energy-momentum tensor, and
\be\label{Veff1}
V_{\text {eff}}(\phi) \equiv V(\phi)-T(x,\phi)
\ee
is the effective potential. Note that, for a negligibly self-gravitating body, the effective potential reduces to
\be\label{Veff2}
V_{\text {eff}}(\phi) = V(\phi)+{\rho}A(\phi)\,,
\ee
where $\rho$ is the matter density of the local environment of the scalar field.

\subsection{Field equations in the weak-field limit}
We are interested in the energy and momentum carried by the (scalar and tensor) GWs at large distances from the source (e.g.  at the position of the detector). So, the tensor field $g_{\mu\nu}$ and the scalar field $\phi$ can be expanded around the two backgrounds as follows:
\begin{align}\label{perturbations}
g_{\mu\nu}=\eta_{\mu\nu}+h_{\mu\nu}\,,\qquad \phi=\ph+\varphi\,\,,
\end{align}
where $\eta_{\mu\nu}$ is the flat Minkowski background, and $\ph$ is the scalar field VEV (scalar background) which depends on the background matter density. Note that, in this paper we do not consider the effect of the cosmological evolution of the scalar field VEV $\ph$ \footnote{This effect on scalar radiation from compact binary systems was studied by Brax and collaborators \cite{Brax2014p225001}.}, i.e., $\ph$ is regarded as a constant in our case.

The bare potential $V(\phi)$ and the coupling function $A(\phi)$ can be expanded in Taylor's series around the scalar background as follows,
\begin{align}\label{VA_expand}
\begin{split}
V(\phi)
&=\V+V_1\varphi+V_2\varphi^2+V_3\varphi^3+\mathcal{O}(\varphi^4)\,,
\\
A(\phi)
&=\A+A_1\varphi+A_2\varphi^2+A_3\varphi^3+\mathcal{O}(\varphi^4)\,,
\end{split}
\end{align}
where $\A\equiv A(\ph)$ is the coupling function VEV, and $\V\equiv V(\ph)$ is the bare potential VEV which can act as the effective cosmological constant to accelerate the expansion of the late universe \cite{Zhang2016p124003}. The inertial mass $m_a(\phi)$ for a strongly self-gravitating body can be expanded in Taylor's series around the scalar background,
\begin{align}
\label{inertial_mass}
\begin{split}
m_a(\phi)=&m_a\bigg[1+s_a\Big(\frac{\varphi}{\phi_{\rm VEV}}\Big)+\frac{1}{2}s'_a\Big(\frac{\varphi}{\phi_{\rm VEV}}\Big)^2
\\&+O\Big(\frac{\varphi}{\phi_{\rm VEV}}\Big)^3\bigg]\,,
\end{split}
\end{align}
where $m_a\equiv m_a(\ph)$ is the inertial mass at the scalar background, and the ``first and second sensitivities'' $s_a$ and
$s'_a$ are defined by \cite{Alsing2012p64041}
\begin{subequations}
\begin{gather}
\label{sensitivities}
%\begin{split}
s_a\equiv\frac{\partial(\ln m_a)}{\partial(\ln \phi)}\bigg|_{\phi_{\rm VEV}},
\\
s'_a\equiv s_a^2-s_a+\frac{\partial^2(\ln m_a)}{\partial(\ln \phi)^2}\bigg|_{\phi_{\rm VEV}}.
%\end{split}
\end{gather}
\end{subequations}

In the weak-field limit, we define small perturbation $\bar{h}_{\mu\nu}=h_{\mu\nu}-\frac{1}{2}\eta_{\mu\nu}h^\lambda_\lambda$, and impose the Lorentz gauge condition $\partial^{\mu}\bar{h}_{\mu\nu}=0$, then the tensor field equation \eqref{tensor_eom} reduces to
\begin{align}\label{linear_tensor_eq}
\square\bar{h}_{\mu\nu}=-16\pi G\tau_{\mu\nu}\,,
\end{align}
where $\square\equiv\eta^{\mu\nu}\partial_{\mu}\partial_{\nu}$ is the flat-space d'Alembertian, and $\tau_{\mu\nu}=T_{\mu\nu}+{\bar t}_{\mu\nu}$ is the total energy-momentum tensor and satisfies the conservation law $\partial_{\nu}\tau^{\mu\nu}=0$\, because of the Bianchi identity. ${\bar t}_{\mu\nu}$ is the common energy-momentum tensor of the scalar and tensor fields, and can be derived by collecting the quadratic and higher-order terms of the perturbations $h_{\mu\nu}$ and $\varphi$ and neglecting the terms involving $V_n$ which correspond to the effects of dark energy. The dark energy effects on GWs from isolated systems  were studied by Ashtekar and collaborators \cite{Ashtekar2016p51101}. If considering only the quadratic terms, ${\bar t}_{\mu\nu}$ can be decomposed as ${\bar t}_{\mu\nu}=T_{\varphi\mu\nu}+t_{\mu\nu}$ (i.e., $h_{\mu\nu}$ and $\varphi$  are decoupled). The quantity
\begin{align}\label{Tuv_varphi}
T_{\varphi\mu\nu}=\partial_\mu\varphi\partial_\nu\varphi-\frac{1}{2}(\partial\varphi)^2{\eta_{\mu\nu}}
\end{align}
is the energy-momentum tensor of the scalar field (or scalar GWs) in the weak-field limit. The quantity $t_{\mu\nu}$ is the stress-energy tensor of gravitational field up to quadratic order in $h_{\mu\nu}$, defined as in GR \cite{Maggiore2007p}. Performing the transverse-traceless (TT) gauge on $t_{\mu\nu}$, we derive the energy-momentum tensor of the tensor GWs,
\begin{align}\label{tuv_TT}
t_{\mu\nu}^{\rm TT}=\f{1}{32{\pi}G}\partial_\mu h_{ij}^{\rm TT}\partial_\nu h^{ij}_{\rm TT}\,,
\end{align}
where ${h}^{\rm TT}_{ij}$ is the TT part of ${h}_{ij}$\,. This result can also be obtained from the Pauli-Fierz action \cite{Maggiore2007p} by using the Noether's theorem.

In the weak-field limit, using the Lorentz gauge condition $\partial^{\mu}\bar{h}_{\mu\nu}=0$,  the scalar field equation \eqref{scalar_eom} reduces to
\begin{align}\label{linear_scalar_eq}
\left(\square-m^2_s\right)\varphi=-16\pi GS\,,
\end{align}
with the distributional source term
\begin{align}
\label{scalar_source}
\begin{split}
S=&-\frac{1}{16\pi G}\left(-\frac{\partial T}{\partial\varphi}+h^{\mu\nu}\partial_{\mu}\partial_{\nu}\varphi+3V_3\varphi^2\right)
\\&+\mathcal O\left(h^3,h^2\varphi,h\varphi^2,\varphi^3\right)\,,
\end{split}
\end{align}
where $m_s$ is the effective mass of the scalar field in a homogeneous background, defined by \eqref{mass_eff}
\begin{align}
m^2_{s}\equiv \frac{\rm d^2 V_{\rm eff}}{\rm d\phi^2}\bigg|_{\ph}\!\!=2(V_2+\rho_b A_2)\,,
\end{align}
which is a positive and monotonically increasing function of the background matter density $\rho_b$. By considering a plane wave $\varphi\sim e^{ik^\lambda x_\lambda}$ and substituting this into $(\square-m_s^2)\varphi=0$, we obtain the relativistic dispersion relation for the scalar mode,
\begin{align}\label{Scalar_dispersion_relation}
\begin{split}
\omega^2={\mathbf k}^2+m_s^2\,,
\end{split}
\end{align}
where $k^\lambda=(\omega,{\mathbf k})$, and $\omega$ and $\mathbf k$ are the frequency and wave vector of the scalar GWs. From this we can further obtain
 \begin{align}\label{Scalar_group_phase}
\begin{split}
v_{s_g}(\omega)&=\sqrt{1-{m_s^2}/{\omega^2}}\,,\\v_{s_p}(\omega)&=\f{1}{\sqrt{1-{m_s^2}/{\omega^2}}}\,,
\end{split}
\end{align}
which are respectively the group and phase speeds of the massive scalar mode, and satisfy the relation $v_{s_g}v_{s_p}=1$. This result implies that the scalar mode in SMG can be excited only if the frequency (energy) of scalar mode is greater than its mass.

\section{Post-Newtonian solution and EIH equations of motion}\label{section3}

In general, in any theory of gravity (including GR), GWs emission depends not only on the dissipative sector of the theory but also on the conservative sector of the theory, and both sectors in alternative theories of gravity are modified relative to GR. In order to understand the dissipative effects, we should first consider the conservative sector of the theory, and investigate the conservative orbital dynamics for compact binary systems in this section.

\subsection{PN scalar solution and sensitivity}
Now, let us derive the static solution of the scalar field equation \eqref{linear_scalar_eq} within the PN approximation  \cite{Will1993p,Will2014p4}. Using the relations \eqref{Tm_trace} and \eqref{inertial_mass}, in the near zone the source term $S$ \eqref{scalar_source} turns into the PN expression,
\begin{align}
\label{scalar_source1}
\begin{split}
S=&-\frac{1}{16\pi G}\Bigg\{\ph^{-1}\sum_as_am_a\delta^3\big(\mathbf{r}-\mathbf{r}_a(t)\big)\bigg[1-\frac12v_a^2
\\&-\frac12\ac{(2)}{h}_k^k+\f{s'_a}{s_a}\ph^{-1}\ac{(2)}{\varphi}\bigg]+\ac{(2)}{h}_{ij}\partial_i\partial_j\ac{(2)}{\varphi}+3V_3\ac{(2)}{\varphi}^{\,2}\Bigg\}+\mathcal O(v^6)\,.
\end{split}
\end{align}
This expression \eqref{scalar_source1} up to leading PN order (i.e., Newtonian order), from Eq. \eqref{linear_scalar_eq} we obtain the field equation in the near zone,
\begin{align}\label{t_ind_sca_equ}
\left(\nabla^2-m^2_s\right)\ac{(2)}{\varphi}=\ph^{-1}\sum_as_am_a\delta^3\big(\mathbf{r}-\mathbf{r}_a(t)\big)\,,
\end{align}
and the solution is
\begin{align}\label{sen_sca_eq}
\ac{(2)}{\varphi}=-2\f{\Mpl^2}{\ph}\sum_a\frac{Gm_as_a}{r_a}e^{-m_sr_a}\,,
\end{align}
where $r_a=\left|\mathbf{r}-\mathbf{r}_a(t)\right|$\,. Note that, this solution is based on the definition of $m_a(\phi)$ in Eq. (\ref{matter_action}) and the related sensitivity of $s_a$ in Eq. (\ref{inertial_mass}).

In addition, based on the action in Eq. (\ref{action_0}), the scalar solution was also derived by using the method of matching the internal and external solutions in Ref. \cite{Zhang2016p124003}, which is briefly reviewed in Appendix \ref{appendix_a}. In this approach, we obtain the solution of scalar field as follows,
\begin{align}\label{t_ind_sca_sol}
\varphi=\sum_a\varphi_a=-\Mpl\sum_a\frac{Gm_a\ep_a}{r_a}e^{-m_sr_a}+\mathcal O(v^4)\,,
\end{align}
with the $a$-th object's screened parameter (or scalar charge)
\begin{align}\label{epsilon_a}
\ep_a\equiv\frac{\ph-\phi_a}{M_\text{Pl}\Phi_{a}}\,,
\end{align}
where $\Phi_a=Gm_a/R_a$ is the surface gravitational potential of the $a$-th object, and $\phi_a$ is the position of the minimum of $V_{\rm eff}$ inside the $a$-th object.

Comparing the above two solutions \eqref{sen_sca_eq} with \eqref{t_ind_sca_sol}, we obtain the useful relation between sensitivity and screened parameter,
\begin{align}\label{s_a}
s_a=\frac{\ph}{2\Mpl}\ep_a\,.
\end{align}
That is to say, the sensitivity $s_a$ is equivalent to the screened parameter (or scalar charge) $\ep_a$ in SMG theories. From Eq. \eqref{epsilon_a} and Eq. \eqref{s_a} we can observe that the sensitivity of the object is inversely proportional to its surface gravitational potential. Therefore, in SMG theories, for the compact objects (such as white dwarfs and neutron stars), the sensitivity effect is very weak (screening mechanism is very strong), and thus the deviations from GR become small and weak. This is completely different from most alternative theories of gravity without screening mechanisms, which generally predict the large non-GR effects for compact objects. Since in these theories, the sensitivities of the object usually increase as its surface gravitational potential increases \cite{will1989}.

\subsection{PN metric solution}\label{PN_Solution}
We solve the tensor field equations \eqref{tensor_eom} within the PN approximation \cite{Will1993p,Will2014p4} in the near zone, where we can neglect the bare potential $V(\phi)$ corresponding to the dark energy. The detailed derivations  are given in Appendix \ref{appendix_b}, and the results are listed below,
\begin{subequations}\label{PN_metric}
\begin{align}
\begin{split}
g_{00}=&-1+2\sum_a\frac{Gm_a}{r_a}-2\bigg(\sum_{a}\frac{Gm_a}{r_a}\bigg)^2+3\sum_a\frac{Gm_av_a^2}{r_a}
\\&-2\sum_a\sum_{b\ne a}\frac{G^2m_am_b}{r_ar_{ab}}\left(1+\frac{1}{2}\ep_a\ep_be^{-m_sr_{ab}}\right)+\mathcal O(v^6)\,,
\end{split}
\end{align}
\begin{align}
\begin{split}
g_{0j}=&-\frac{7}{2}\sum_a\!\frac{Gm_av_a^j}{r_a}\!-\!\frac {1}{2}\sum_{a}\!\frac{Gm_a}{r_a^3}(\mathbf{r}_a\!\cdot\! \mathbf{v}_a)(r^j\!-\!r_a^j)\!+\!\mathcal O(v^5)\,,
\end{split}
\end{align}
\begin{align}
\begin{split}
g_{ij}=&\delta_{ij}\left(1+2\sum_a\frac{Gm_a}{r_a}\right)+\mathcal O(v^4)\,,
\end{split}
\end{align}
\end{subequations}
where $m_a$ is the inertial mass of the $a$-th object, $\ep_a$ is its screened parameter, $v_a$ is its velocity, $m_s$ is the effective mass of the scalar, $r_a=\left|\mathbf{r}-\mathbf{r}_a(t)\right|$,  and $r_{ab}=\left|\mathbf{r}_a(t)-\mathbf{r}_b(t)\right|$\,. Obviously, the above results can reduce to the GR case in the limit where every object's screened parameter $\ep_a\rightarrow0$\,.

Substituting these PN solutions \eqref{t_ind_sca_sol} and \eqref{PN_metric} into the  source term $S$ \eqref{scalar_source1}, and using Eq. \eqref{t_ind_sca_equ}, we obtain the PN expression of the  source term $S$ in the near zone,
\begin{align}
\label{scalar_source2}
\begin{split}
S=&-\frac{\Mpl}{4}\sum_a\ep_am_a\delta^3\big(\mathbf{r}-\mathbf{r}_a(t)\big)\bigg[1-\frac12v_a^2-\sum_{b\ne a}\frac{Gm_b}{r_b}
\\&-\frac{s'_a}{s_a}\frac{\Mpl}{\ph}\times\sum_{b\ne a}\frac{Gm_b\ep_b}{r_b}e^{-m_sr_b}+\mathcal O(v^4)\bigg]\,,
\end{split}
\end{align}
where we have neglected the terms involving $V_n$ which correspond to the effects of dark energy, since these effects are very weak in the near zone.

\subsection{Violation of the WEP and \\EIH equations of motion}
The weak equivalence principle (WEP) is defined as the universality of free fall for bodies. We know that the WEP is satisfied in GR where the sensitivities are absent. However, the WEP generally does not hold in alternative theories of gravity where the sensitivities are not zero in general. This is because that the sensitivities characterize how the properties (e.g.  mass) of a compact object change with its motion relative to the additional field of the theory. Therefore, different bodies respond differently to motion relative to the ambient field, and thus move along different trajectories. Thus, the WEP is violated in the theories \cite{DiCasola2015p39}. In other words, the violation of the WEP is due to the additional field force (fifth force), which depends on the properties (besides mass, e.g.  self-gravitational binding energy) of the object.

In SMG, the first sensitivity is equivalent to the screened parameter, which affects both the conservative and dissipative sectors of theory. For the former one, the screened parameter modifies the conservative orbital dynamics of compact systems, which can be derived from the matter action \eqref{matter_action} by using the method of Einstein, Infeld and Hoffmann (EIH) \cite{Einstein1938p65}. Using the expansion of $m_a(\phi)$ in \eqref{inertial_mass} and the PN expressions of the scalar and tensor fields in \eqref{t_ind_sca_sol} and \eqref{PN_metric}, from the matter action \eqref{matter_action} we obtain the EIH Lagrangian up to Newtonian order,
\begin{align}
\begin{split}
\label{LEIH}
L_{\rm EIH}&\!=-\!\sum_am_a(\phi)\frac{d\tau_a}{dt}
\\&\!=-\!\sum_am_a\!\left(\!1-\frac12v_a^2\right)\!+\frac12\sum_{a}\!\sum_{b\ne a}\frac{\mathcal G_{ab}m_am_b}{r_{ab}}+\!\mathcal O(v^4),
\end{split}
\end{align}
with the effective gravitational `constant'
\begin{align}\label{G_ab1}
\mathcal G_{ab}\equiv G\left(1+\frac12\ep_a\ep_be^{-m_sr_{ab}}\right)\,.
\end{align}
Note that, this result is manifestly symmetric under interchange of all pairs of particles. 

Substituting the EIH Lagrangian into the Euler-Lagrange equation yields the $n$-body equations of motion up to Newtonian order,
\begin{align}
\begin{split}
\label{a_a}
\mathbf{a}_{a}&=-\sum_{b\ne a}\!\frac{{\mathscr G}_{ab}m_b}{r^2_{ab}}\mathbf{\hat{r}}_{ab}\,,
\end{split}
\end{align}
with
\begin{align}\label{G_ab2}
{\mathscr G}_{ab}\equiv G\left[1+\frac12\ep_a\ep_b(1+m_sr_{ab})e^{-m_sr_{ab}}\right]\,,
\end{align}
where $\mathbf{a}_{a}\equiv d^2\mathbf{r}_{a}/dt^2$ is the acceleration of the $a$-th object, $\mathbf{\hat{r}}_{ab}$ is the unit direction vector from the $b$-th object to the $a$-th object, and $r_{ab}=\left|\mathbf{r}_a(t)-\mathbf{r}_b(t)\right|$\,. Note that, the Yukawa-like terms involving the screened parameters violate the WEP and the gravitational inverse-square law. In the near zone, the separation $r_{ab}$ is always much less than the Compton wavelength $m^{-1}_s$ (which roughly is cosmological scales), i.e., $m_s r_{ab}\ll1$ is satisfied. Using this relation, both the expressions \eqref{G_ab1} and \eqref{G_ab2} reduce to
\begin{align}\label{G_ab3}
\mathcal G_{ab}={\mathscr G}_{ab}= G\left(1+\frac12\ep_a\ep_b\right)\,.
\end{align}
Note that, this result satisfies the inverse-square law but still violates the WEP, since the screened parameters (or scalar charges) of different bodies are different.

Now let us consider a binary system of compact objects. The most well-known dissipative effect is the orbital period decay due to the emission of gravitational radiation. In fact, it was the monitoring of the orbital period that led to the first indirect detection of GWs by Hulse and Taylor \cite{Taylor1994p711}. Because the orbital motion satisfies the inverse-square law in Eq. \eqref{a_a} and \eqref{G_ab3}, the orbital period decay rate $\dot{P}$ can be written as
\begin{align}\label{P_E}
\f{\dot{P}}{P}=-\f{3}{2}\f{\dot{E}}{E}\,, 
\end{align}
where the orbital period $P$ satisfies the Kepler's third law
\begin{align}\label{Kepler_3rd}
(2\pi/P)^2a^3=\mathcal Gm\,,
\end{align}
and
\begin{align}\label{binding_energy}
E=-\f{\mathcal Gm\mu}{2a}
\end{align}
is the orbital binding energy of the system.
Here, 
\begin{align}\label{G_12}
\mathcal G\equiv\mathcal G_{12}=G\left(1+\frac12\ep_1\ep_2\right)
\end{align}
is the effective gravitational coupling constant between two compact objects (labeled by \( 1 \) and \( 2 \)), $a$ is the semimajor axis,  and $m\equiv m_1+m_2$, $\mu\equiv m_1m_2/m$ are the total and reduced masses of the system. Note that, the inverse-square law guarantees that these relations \eqref{P_E}, \eqref{Kepler_3rd} and \eqref{binding_energy} hold in SMG theories. From the relation in Eq. (\ref{P_E}), we find that the orbital decay of the binary system is directly determined by the energy loss of the system, which will be addressed in the next section.

\section{Gravitational radiation from compact binaries}\label{section4}
In GR, we know that the leading order energy flux is quadrupole radiation flux. However, besides quadrupole radiation, a general scalar-tensor theory also predicts monopole and dipole radiations \cite{Eardley1975p59, Will1977p826}. In this section, we focus on the dissipative effects of SMG, calculate the rate of the energy loss due to the emission of tensor and scalar gravitational radiations (including monopole, dipole, quadrupole, and dipole-octupole radiations), and derive their contributions to the change in the orbital period.

\subsection{Tensor and scalar energy fluxes}
The energy flux of GWs is defined as the energy of GWs flow per unit time at a large distance from the source. Since the total energy of the system is a conserved quantity, the rate of change of the orbital binding energy $\dot{E}$ is equal to minus the total energy flux $\mathcal F$ carried away from compact binary system by GWs, i.e.,
\begin{align}\label{E_F}
\dot{E}=-\mathcal F\,.
\end{align}

In GR, the energy flux is only due to the propagation of tensor mode, but in a general scalar-tensor theory, gravitational radiation comes from both scalar and tensor modes. In addition, in gravity theories with vector fields like TeVeS theory \cite{Bekenstein2004p83509, Sagi2010p64031} and Einstein-{\ae}ther theory \cite{Yagi2014p161101, Yagi2014p84067}, vector modes also exist. The energy flux carried by all propagating degrees of freedom can be derived directly from the Lagrangian of the theory by investigating the Noether charges and currents in the theory. Here, we will derive the formulae to calculate the tensor and scalar energy fluxes in the general SMG.

In the wave zone (far zone), because of the absence of matter energy-momentum tensor $T_{\mu\nu}$, we have the conservation law $\partial_{\nu}\left(t_{\rm TT}^{\mu\nu}+T_{\varphi}^{\mu\nu}\right)=0$\,. Since $h_{\mu\nu}$ and $\varphi$ are decoupled, the energy-momentum tensors (i.e., Noether currents) of the tensor and scalar GWs are respectively conserved, i.e., $\partial_{\nu}t_{\rm TT}^{\mu\nu}=0$ and $\partial_{\nu}T_{\varphi}^{\mu\nu}=0$\,. Thus, we can investigate them separately.

According to the conservation law $\partial_{\nu}t_{\rm TT}^{\mu\nu}=0$, from the energy-momentum tensor of the tensor GWs \eqref{tuv_TT}, we obtain the tensor energy flux 
\begin{align}\label{tensor_flux1}
\begin{split}
{\mathcal F}_g&=r^2\int \!d\Omega\left\langle{t^{0r}_{\rm TT}}\right\rangle
\\&=-\frac{r^2}{32\pi G}\int \!d\Omega\left\langle\partial_0{h}^{\rm TT}_{ij}\partial_r{h}^{\rm TT}_{ij}\right\rangle,
\end{split}
\end{align}
where the angular brackets represent a time average over a period of the system's motion, ${h}^{\rm TT}_{ij}$ is the TT part of ${h}_{ij}$, and $\Omega$ is the solid angle. The massless tensor mode propagates with the speed of light, and $h_{ij}^{\rm TT}(t,{\bf r})$ takes the form $({1}/{r})f_{ij}(t-r)$, so we have $\partial_r{h}^{\rm TT}_{ij}=-\partial_0{h}^{\rm TT}_{ij}+\mathcal O(1/r^2)$ at large distances. Using this, the tensor energy flux \eqref{tensor_flux1} can be further simplified to
\begin{align}\label{tensor_flux2}
\begin{split}
{\mathcal F}_g=\frac{r^2}{32\pi G}\int \!d\Omega\left\langle\partial_0{h}^{\rm TT}_{ij}\partial_0{h}^{\rm TT}_{ij}\right\rangle.
\end{split}
\end{align}
This expression is exactly the same as that in GR.

The scalar energy flux can be derived from the energy-momentum tensor of the scalar GWs \eqref{Tuv_varphi} by using the conservation law $\partial_{\nu}T_{\varphi}^{\mu\nu}=0$,
\begin{align}\label{Scalar_flux}
\begin{split}
{\mathcal F}_{\phi}&=r^2\int \!d\Omega\left\langle{T_{\varphi}^{0r}}\right\rangle
\\&=-r^2\int \!d\Omega\left\langle\partial_0{\varphi}\partial_r{\varphi}\right\rangle.
\end{split}
\end{align}
Unlike Eq. \eqref{tensor_flux1}, this expression \eqref{Scalar_flux} cannot be further simplified, since the speed of propagation of the massive scalar mode changes with its frequency (see Eq. \eqref{Scalar_group_phase}).

\subsection{Tensor radiation}

By using a retarded Green's function, performing the time integral, we obtain the formal solution of the linearized tensor wave equation \eqref{linear_tensor_eq},
\begin{align}
\label{weak_tensor_general_sol}
\bar{h}^{\mu\nu}(t,\mathbf{r})=4G\int_\mathcal{N}\!d^3\mathbf{r'}\,\f{\tau^{\mu\nu}(t\!-\!|\mathbf{r\!-\!r'}|,\,\mathbf{r'})}{|\mathbf{r\!-\!r'}|}\,.
\end{align}
Here, the spatial (source point $\mathbf{r'}$) integration region $\mathcal{N}$ is over the near zone, the field point $\mathbf{r}$ is in the wave zone (far zone), such that $|\mathbf{r}'|\ll|\mathbf{r}|$. Considering this condition and making the slow-motion approximation, we can expand the integrand in powers of ($\mathbf{n\cdot r'}$)\, as follows,
\begin{align}
\bar{h}^{\mu\nu}(t,\mathbf{r})=\f{4G}{r}\sum_{{\ell}=0}^\infty\f{1}{{\ell}!}\f{\partial^{\ell}}{\partial t^{\ell}}\!\int_\mathcal{N}\!\tau^{\mu\nu}(t\!-\!r,\mathbf{r'})(\mathbf{n\cdot r'})^{\ell}d^3\mathbf{r'}\,,
\end{align}
where $\mathbf{n}=\mathbf{r}/r$ is the unit vector in the $\mathbf{r}$ direction. Because of the conservation law $\partial_{\nu}\tau^{\mu\nu}=0$, the spatial components $\bar{h}^{ij}$ up to leading order ($\ell=0$), can be rewritten as
\begin{align}\label{tau_multipole}
\begin{split}
\bar{h}^{ij}(t,\mathbf{r})&=\f{4G}{r}\int\tau^{ij}(t-r,\mathbf{r'})d^3\mathbf{r'}
\\&=\f{2G}{r}\f{\partial^2}{\partial t^2}\int\tau^{00}(t\!-\!r,\mathbf{r'})\,r'^{i}r'^{j}d^3\mathbf{r'}\,,
\end{split}
\end{align}
which only involves the quadrupole moment of $\tau^{00}$, like in GR, there is neither monopole nor dipole radiations in tensor gravitational radiation. This emerges as a consequence of the fact that the tensor graviton is a massless spin-2 particle \cite{Maggiore2007p}. The quantity $\tau^{00}$ is the total energy density of both matter and (scalar and tensor) fields. Note that at the leading PN order, the fields energy density is negligible, so from Eq. \eqref{Tuv_matter} we obtain the expression of $\tau^{00}$ as follows,
\begin{align}\label{tau_00}
\begin{split}
\tau^{00}(t,\mathbf r)=\sum_am_a\delta^3(\mathbf{r}-\mathbf{r}_a(t))\,.
\end{split}
\end{align}
Substituting this into Eq. \eqref{tau_multipole} yields
\begin{align}\label{h_ij_M}
\bar{h}^{ij}(t,\mathbf r)=\f{2G}{r}\f{d^2}{d t^2}M^{ij}\bigg|_{\rm ret}\,,
\end{align}
with the mass quadrupole moment 
\begin{align}\label{mass_quadrupole}
M^{ij}(t)=\sum_a m_a r_a^i(t) r_a^j(t)\,,
\end{align}
where the subscript `ret' means that the quantity $M^{ij}$ is evaluated at the retarded time $t-r$\,. The TT part of ${h}^{ij}$ is ${h}^{ij}_{\rm TT}=\Lambda_{ij,kl}{h}^{kl}=\Lambda_{ij,kl}{\bar{h}}^{kl}$, where the projector $\Lambda_{ij,kl}$ is the Lambda tensor as defined in \cite{Maggiore2007p}. Using Eqs. \eqref{h_ij_M} and \eqref{mass_quadrupole}, from Eq. \eqref{tensor_flux2} we obtain the tensor quadrupole flux
\begin{align}\label{Q_F_g}
\begin{split}
{\mathcal F}_g^{Q}=\f{G}{5}\left\langle\dddot{M}^{kl}\dddot{M}^{kl}-\f{1}{3}\left(\dddot{M}^{kk}\right)^2\right\rangle\,,
\end{split}
\end{align}
where we have performed the integral over the solid angle. The overdots denote derivatives with respect to coordinate time, and the angular brackets represent a time average over an orbital period.  At leading PN order, the tensor quadrupole flux \eqref{Q_F_g} in SMG behaves as in GR.

Now let us consider a compact binary  (labeled by 1 and 2) with quasi-circular orbit, which is parameterized in the center of mass frame by
\begin{align}\label{x_y_z}
\begin{split}
x_1(t)&=-R_1\cos(\om t),\,y_1(t)=-R_1\sin(\om t),\,z_1=0,
\\
x_2(t)&=R_2\cos(\om t),~~~y_2(t)=R_2\sin(\om t),~~~z_2=0,
\end{split}
\end{align}
where $\om$ is the orbital frequency, and $R_1$ and $R_2$ are the orbital radiuses of the two components of the binary system. Substituting these into Eq. \eqref{mass_quadrupole}, using the Kepler's third law \eqref{Kepler_3rd}, from Eq. \eqref{Q_F_g} we obtain \begin{align}\label{quadrupole_radiation_flux}
\begin{split}
{\mathcal F}_g^{Q}=\frac{32G\mu^2(\mathcal Gm)^3}{5R^5}\,,
\end{split}
\end{align}
where $R=R_1+R_2$ is the separation between the two components of the system, and $\mathcal G=G\left(1+\frac12\ep_1\ep_2\right)$ is the effective gravitational coupling constant between the two components.

\subsection{Scalar radiation}

Now, let us turn to the dissipative effects of the scalar sector of SMG, and show that there are monopole, dipole, and dipole-octupole radiations in the scalar sector, besides quadrupole radiation.

The massive scalar wave equation \eqref{linear_scalar_eq} can be solved by using Green's function method,
\begin{align}\label{scalar_Green}
\left(\square-m^2_s\right)G(x,x')=-4\pi \delta^4(x-x')\,,
\end{align}
and the formal solution of Eq. \eqref{linear_scalar_eq} is
\begin{align}\label{scalar_formal_solution}
\varphi(x)=4G\int d^4x'S(x')G(x,x')\,.
\end{align}
The Green's function in Eq. \eqref{scalar_Green} is given by \cite{Poisson2011p7,M.Morse1953p}
\begin{align}
\begin{split}
G(x,x')=& \frac{\delta(t\!-\!t'\!-\!|\mathbf{r\!-\!r'}|)}{|\mathbf{r\!-\!r'}|}\!-\!\Theta(t\!-\!t'\!-\!|\mathbf{r\!-\!r'}|)
\\&
\times\frac{m_sJ_1\!\big(m_s\!\sqrt{(t\!-\!t')^2\!-\!|\mathbf{r\!-\!r'}|^2}\big)}{\sqrt{(t\!-\!t')^2\!-\!|\mathbf{r\!-\!r'}|^2}}\,,
\end{split}
\end{align}
where $\delta$ is the Dirac delta-function, $\Theta$ is the Heaviside function, and $J_1$ is the Bessel function of the first kind. Substituting this into \eqref{scalar_formal_solution} and performing the time $t'$ integral, we obtain the formal solution
\begin{align}\label{t_dep_sca_sol1}
\begin{split}
\varphi(t,\mathbf{r})=\,&4G\!\int_{0}^{\infty}\!\!\!\!dzJ_1(z)\!\!\int_{\!\mathcal{N}}\!\! d^3\mathbf{r'}\Bigg\{\f{S\big(t\!-\!|\mathbf{r\!-\!r'}|,\,\mathbf{r'}\big)}{|\mathbf{r\!-\!r'}|}
\\&~~~-\f{S\big(t\!-\!\sqrt{|\mathbf{r\!-\!r'}|^2\!+\!({z}/{m_s})^2},\,\mathbf{r'}\big)}{\sqrt{|\mathbf{r\!-\!r'}|^2\!+\!({z}/{m_s})^2}}\Bigg\}\,,
\end{split}
\end{align}
where we have used the identity $\int_{0}^{\infty}\!J_1(z)dz=1$ and made the substitution $z=m_s\sqrt{(t-t')^2-|\mathbf{r-r'}|^2}$\,. Here, the spatial (source point $\mathbf{r'}$) integration region $\mathcal{N}$ is over the near zone, the field point $\mathbf{r}$ is in the wave zone (far zone), such that $|\mathbf{r}'|\ll|\mathbf{r}|$, and considering the slow-motion approximation, the integrand in Eq. \eqref{t_dep_sca_sol1} can be expanded in Taylor's series of $(\mathbf{n \cdot r'})$,
\begin{align}\label{t_dep_sca_sol2}
\begin{split}
\varphi(t,\mathbf{r})=\,&\f{4G}{r}\int_{0}^{\infty}\!dzJ_1(z)\sum_{{\ell}=0}^\infty\f{1}{{\ell}!}\f{\partial^{\ell}}{\partial t^{\ell}}\int_{\mathcal{N}}d^3\mathbf{r'}(\mathbf{n \cdot r'})^{\ell} 
\\&~~~~\times\bigg\{S\big(t\!-\!r,\,\mathbf{r'}\big)-\f{S\big(t\!-\!{r}{u(r,z)},\,\mathbf{r'}\big)}{u^{\ell+1}\!(r,z)}\bigg\},
\end{split}
\end{align}
with
\begin{align}
\begin{split}
u(r,z)\equiv{\sqrt{1+\Big(\f{z}{m_sr}\Big)^2}}\,,
\end{split}
\end{align}
where $\mathbf{n}=\mathbf{r}/r$ is the unit vector in the $\mathbf{r}$ direction.

Substituting the source term $S$ \eqref{scalar_source2} into this formal solution \eqref{t_dep_sca_sol2} and performing the spatial $\mathbf{r'}$ integral, and we have
\begin{align}\label{t_dep_sca_sol3}
\begin{split}
\varphi(t,\mathbf{r})=&-\Mpl\f{G}{r}\!\int_{0}^{\infty}\!\!dzJ_1(z)\sum_{{\ell}=0}^\infty\f{1}{\ell!}n_{L}{\partial^{\ell}_t}\M_{\ell}^L\,,
\end{split}
\end{align}
with the scalar multipole moments
\begin{align}\label{mass_moment_0}
\begin{split}
\M_{\ell}^L\equiv&\M_{\ell}^{i_1i_2\cdots i_{\ell}}(t,r,z)
\\=&\sum_a\ep_a\bigg[M_a(t\!-\!r)\cdot r_a^L(t\!-\!r)-u^{-(\ell+1)}(r,z) 
\\&\quad~~\times M_a(t\!-\!{r}{u(r,z)})\cdot r_a^L(t\!-\!{r}{u(r,z)})\bigg]\,,
\end{split}
\end{align}
and the mass
\begin{align}\label{t_dep_mass}
\begin{split}
M_a(t)\equiv& m_a\bigg[1-\f12v_a^2(t)-\sum_{b\ne a}\f{Gm_b}{r_{ab}(t)}
\\&~~~~-\f{s'_a}{s_a}\f{\Mpl}{\ph}\sum_{b\ne a}\f{Gm_b\ep_b}{r_{ab}(t)}e^{-m_sr_{ab}(t)}\bigg]\,,
\end{split}
\end{align}
where the quantities $n_L$ and $r_a^L(t)$ are defined by 
\begin{align}
\begin{split}
n_L\!\equiv n_{i_1}n_{i_2}\!\cdots\! n_{i_\ell},\quad~ r_a^L(t)\equiv r_a^{i_1}(t) r_a^{i_2}(t)\!\cdots\!  r_a^{i_{\ell}}(t)\,.
\end{split}
\end{align}
Taking the spatial gradient of the scalar field \eqref{t_dep_sca_sol3} and neglecting the higher order terms $\mathcal O(1/r^2)$, we obtain  
\begin{align}\label{part_r_scalar}
\begin{split}
\partial_r\varphi(t,\mathbf{r})=&\Mpl\f{G}{r}\!\int_{0}^{\infty}\!\!dzJ_1(z)\sum_{{\ell}=0}^\infty\f{1}{\ell!}n_{L}{\partial^{\ell+1}_t}\M_{\ell+1}^L\,,
\end{split}
\end{align}
where we again define a new scalar multipole moments
\begin{align}\label{mass_moment_1}
\begin{split}
\M_{\ell+1}^L\equiv&\M_{\ell+1}^{i_1i_2\cdots i_{\ell}}(t,r,z)
\\=&\sum_a\ep_a\bigg[M_a(t\!-\!r)\cdot r_a^L(t\!-\!r)-u^{-(\ell+2)}(r,z) 
\\&\quad~~\times M_a(t\!-\!{r}{u(r,z)})\cdot r_a^L(t\!-\!{r}{u(r,z)})\bigg]\,.
\end{split}
\end{align}
From Eqs. \eqref{mass_moment_0} and \eqref{mass_moment_1} we find that all scalar multipole moments are  suppressed by the screened parameters of the objects, since the scalar gravitational charge is the screened parameter.

Substituting Eqs. \eqref{t_dep_sca_sol3} and \eqref{part_r_scalar} into Eq. \eqref{Scalar_flux} and performing the integral over the solid angle, we obtain the scalar energy flux %% up to quadrupole radiation ($\ell=2$),
\begin{align}
\label{scalar_quadrupole_flux}
\begin{split}
{\mathcal F}_{\phi}=&\frac{G}{2}\iint dz_1dz_2J_1(z_1)J_1(z_2)\bigg{\langle}\dot{\M}_0\dot{\M}_1
\\&+\frac{1}{6}\Big(2\ddot{\M}_1^k\ddot{\M}_2^k+\dot{\M}_0\dddot{\M}_3^{kk}+\dot{\M}_1\dddot{\M}_2^{kk}\Big)
\\&+\frac{1}{60}\Big(2\dddot{\M}_2^{kl}\dddot{\M}_3^{kl}+\dddot{\M}_2^{kk}\dddot{\M}_3^{ll}\Big)
\\&+\frac{1}{30}\Big(\ddot{\M}_1^{k}\ddddot{\M}_4^{kll}+\ddot{\M}_2^{k}\ddddot{\M}_3^{kll}\Big)\bigg{\rangle}\,,
\end{split}
\end{align}
%%%%%%%%%%%%
%%%%%%%%%%%%
\begin{comment}
\begin{align}
\label{scalar_quadrupole_flux}
\begin{split}
{\mathcal F}_{\phi}=&\frac{G}{2}\iint dz_1dz_2J_1(z_1)J_1(z_2)\bigg{\langle}\dot{\M}_0\dot{\M}_1
\\&+\frac{1}{6}\Big(2\ddot{\M}_1^k\ddot{\M}_2^k+\dot{\M}_0\dddot{\M}_3^{kk}+\dot{\M}_1\dddot{\M}_2^{kk}\Big)
\\&+\frac{1}{120}\Big(4\dddot{\M}_2^{kl}\dddot{\M}_3^{kl}+2\dddot{\M}_2^{kk}\dddot{\M}_3^{ll}+4\ddot{\M}_1^{k}\ddddot{\M}_4^{kll}
\\&+4\ddot{\M}_2^{k}\ddddot{\M}_3^{kll}+\dot{\M}_0{\partial_t^5}{\M}_5^{kkll}+\dot{\M}_1{\partial_t^5}{\M}_4^{kkll}\Big)\bigg{\rangle}\,,
\end{split}
\end{align}
\end{comment}
%%%%%%%%%%%%
%%%%%%%%%%%%
where the angular brackets represent a time average over a period of the system's motion, the overdots denote derivatives with respect to coordinate time, and we have used the identity \cite{Maggiore2007p}
\begin{align}
\label{unit_int_identity}
\int\!\f{d\Omega}{4\pi} n_{i_1}n_{i_2}\!\cdots\! n_{i_{k}}\!=\!\left\{
\begin{matrix}\!\!
0\qquad\qquad\qquad~~~&{\rm for}~~k={\rm odd} 
\\[0.6 em]
\f{\delta_{i_1i_2}\delta_{i_3i_4}\cdots\delta_{i_{k\!-\!1}i_{k}}+\cdots}{(k\,+\,1)!!}~&{\rm for}~~k={\rm even}
\end{matrix}
\right.,
\end{align}
where the final dots denote all possible pairing of indices.

Now let us specialize our calculations to a compact binary with quasi-circular orbit parameterized in the center of mass frame by Eq. \eqref{x_y_z}. Substituting Eq. \eqref{x_y_z} into the scalar multipole moments \eqref{mass_moment_0} and \eqref{mass_moment_1}, we obtain the time derivatives of monopole, dipole, quadrupole, and octupole moments as follows:

\emph{1. Monopole:}
\begin{align}\label{Monopole_circular}
\begin{split}
\dot{\M}_0=\dot{\M}_1=0\,.
\end{split}
\end{align}

\emph{2. Dipole:}
\begin{subequations}
\begin{align}
\begin{split}
\ddot{\M}_1^k\!=&\!-{\Big(\E_d-\f{G\mu}{2R}{\bar{\E}_d}\Big)}\mu\omega^2\!R
\\&\times\Big[\cos(\omega (t\!-\!r))-u^{-2}\!\cos(\omega (t\!-\!ru)),
\\&~~~~~\sin(\omega (t\!-\!r))-u^{-2}\sin(\omega (t\!-\!ru)),\quad0\Big]\,,
\end{split}
\end{align}
\begin{align}
\begin{split}
\ddot{\M}_2^k\!=&\!-{\Big(\E_d-\f{G\mu}{2R}{\bar{\E}_d}\Big)}\mu\omega^2\!R
\\&\times\Big[\cos(\omega (t\!-\!r))-u^{-3}\!\cos(\omega (t\!-\!ru)),
\\&~~~~~\sin(\omega (t\!-\!r))-u^{-3}\sin(\omega (t\!-\!ru)),\quad0\Big]\,.
\end{split}
\end{align}
\end{subequations}

\emph{3. Quadrupole:}
\begin{equation}\label{polarization matrices}
   \begin{array}{cc}
      \dddot{\M}_2^{kl}\!= \!\left(\!\!
        \begin{array}{ccc}
          \dddot{\M}_2^{11} & \dddot{\M}_2^{12} & 0 \\
          \dddot{\M}_2^{12} & -\dddot{\M}_2^{11} & 0 \\
          0 & 0 & 0
        \end{array}
      \!\right)\!,~~
     \dddot{\M}_3^{kl}\!= \!\left(\!\!
        \begin{array}{ccc}
         \dddot{\M}_3^{11} & \dddot{\M}_3^{12} & 0 \\
          \dddot{\M}_3^{12} & -\dddot{\M}_3^{11} & 0 \\
          0 & 0 & 0
        \end{array}
      \!\right)\!,
   \end{array}
\end{equation}
with the components
\begin{subequations}
\begin{align}
\begin{split}
\dddot{\M}_2^{11}&\!=\!4\E_q\mu\omega^3\!R^2\Big[\sin(2\omega (t\!-\!r))-u^{-3}\!\sin(2\omega (t\!-\!ru))\Big]\,,
\end{split}
\end{align}
\begin{align}
\begin{split}
\dddot{\M}_2^{12}&\!=\!-4\E_q\mu\omega^3\!R^2\Big[\cos(2\omega (t\!-\!r))-u^{-3}\!\cos(2\omega (t\!-\!ru))\Big]\,,
\end{split}
\end{align}
\begin{align}
\begin{split}
\dddot{\M}_3^{11}&\!=\!4\E_q\mu\omega^3\!R^2\Big[\sin(2\omega (t\!-\!r))-u^{-4}\!\sin(2\omega (t\!-\!ru))\Big]\,,
\end{split}
\end{align}
\begin{align}
\begin{split}
\dddot{\M}_3^{12}&\!=\!-4\E_q\mu\omega^3\!R^2\Big[\cos(2\omega (t\!-\!r))-u^{-4}\!\cos(2\omega (t\!-\!ru))\Big]\,.
\end{split}
\end{align}
\end{subequations}

\emph{4. Octupole:}
\begin{subequations}
\begin{align}
\begin{split}
\ddddot{\M}_3^{1kk}=&\E_o\mu\om^4R^3[\cos(\om(t-r))-u^{-4}\cos(\om(t-ru))]\,,
\end{split}
\end{align}
\begin{align}
\begin{split}
\ddddot{\M}_3^{2kk}=&\E_o\mu\om^4R^3[\sin(\om(t-r))-u^{-4}\sin(\om(t-ru))]\,,
\end{split}
\end{align}
\begin{align}
\begin{split}
\ddddot{\M}_4^{1kk}=&\E_o\mu\om^4R^3[\cos(\om(t-r))-u^{-5}\cos(\om(t-ru))]\,,
\end{split}
\end{align}
\begin{align}
\begin{split}
\ddddot{\M}_4^{2kk}=&\E_o\mu\om^4R^3[\sin(\om(t-r))-u^{-5}\sin(\om(t-ru))]\,,
\end{split}
\end{align}
\end{subequations}
where the dummy indices just indicate summation.
Here, $\om$ is the orbital frequency, $\mu$ is the reduced mass of the system, $R$ is the separation between the two components of the system, $u=\sqrt{1+z^2/(m_sr)^2}$, and we have defined 
\begin{subequations}
\begin{align}
\begin{split}
\E_d\equiv\ep_2-\ep_1\,,
\end{split}
\end{align}
\begin{align}
\begin{split}
{\bar{\E}_d}\equiv2(\ep_2-\ep_1)+3\left(\f{\ep_2m_1}{m_2}-\f{\ep_1m_2}{m_1}\right)\,,
\end{split}
\end{align}
\begin{align}
\begin{split}
\E_q\equiv\f{\ep_2m_1+\ep_1m_2}{m_1+m_2}\,,
\end{split}
\end{align}
\begin{align}
\begin{split}
\E_o\equiv\f{\ep_2m_1^2-\ep_1m_2^2}{(m_1+m_2)^2}\,,
\end{split}
\end{align}
\end{subequations}
where the subscripts $d$, $q$, and $o$ denote dipole, quadrupole, and octupole, respectively. Note that, from Eq. \eqref{Monopole_circular} we can observe that there is no monopole radiation contribution to leading order in the quasi-circular orbit case.

Using the above results, the scalar energy flux \eqref{scalar_quadrupole_flux} can be further simplified to
\begin{align}
\begin{split}
{\mathcal F}_{\phi}={\mathcal F}_{\phi}^{D}+{\mathcal F}_{\phi}^{Q}+{\mathcal F}_{\phi}^{DO}\,,
\end{split}
\end{align}
with the scalar dipole flux
\begin{subequations}\label{D_Q_O_scalar_flux}
\begin{align}
\begin{split}\label{Dipole_scalar_flux_1}
{\mathcal F}_{\phi}^{D}=&\frac{G}{6}\iint dz_1dz_2J_1(z_1)J_1(z_2)\ddot{\M}_1^k(z_1)\ddot{\M}_2^k(z_2)
\\=&\frac{G(\mathcal Gm\mu)^2}{6R^4}{\Big(\E_d^2-\f{G\mu}{R}{\E_d\bar{\E}_d}\Big)}\bigg\{1
\\&-\cos(\omega r)\langle\cos(\omega ru)\rangle_2-\sin(\omega r)\langle\sin(\omega ru)\rangle_2
\\&-\left(\cos(\omega r)-\langle\cos(\omega ru)\rangle_2\right)\langle\cos(\omega ru)\rangle_3
\\&-\left(\sin(\omega r)-\langle\sin(\omega ru)\rangle_2\right)\langle\sin(\omega ru)\rangle_3\bigg\}\,,
\end{split}
\end{align}
the scalar quadrupole flux
\begin{align}\label{Quadrupole_scalar_flux_1}
\begin{split}
{\mathcal F}_{\phi}^{Q}=&\frac{G}{60}\iint dz_1dz_2J_1(z_1)J_1(z_2)\dddot{\M}_2^{kl}(z_1)\dddot{\M}_3^{kl}(z_2)
\\=&\frac{8G\mu^2(\mathcal Gm)^3}{15R^5}\E_q^2\bigg\{1-\cos(2\omega r)\langle\cos(2\omega ru)\rangle_3
\\&-\sin(2\omega r)\langle\sin(2\omega ru)\rangle_3
\\&-\left(\cos(2\omega r)-\langle\cos(2\omega ru)\rangle_3\right)\langle\cos(2\omega ru)\rangle_4
\\&-\left(\sin(2\omega r)-\langle\sin(2\omega ru)\rangle_3\right)\langle\sin(2\omega ru)\rangle_4\bigg\}\,,
\end{split}
\end{align}
and the scalar dipole-octupole flux
\begin{widetext}
\begin{align}\label{DO_scalar_flux}
\begin{split}
{\mathcal F}_{\phi}^{DO}=&\frac{G}{60}\iint dz_1dz_2J_1(z_1)J_1(z_2)\Big[\ddot{\M}_1^{k}(z_1)\ddddot{\M}_4^{kll}(z_2)+\ddot{\M}_2^{k}(z_1)\ddddot{\M}_3^{kll}(z_2)\Big]
\\=&-\frac{G\mu^2(\mathcal Gm)^3}{60R^5}\E_d\E_o\bigg\{2-\cos(\omega r)\Big(\langle\cos(\omega ru)\rangle_2+\langle\cos(\omega ru)\rangle_3+\langle\cos(\omega ru)\rangle_4+\langle\cos(\omega ru)\rangle_5\Big)
\\&-\sin(\omega r)\Big(\langle\sin(\omega ru)\rangle_2+\langle\sin(\omega ru)\rangle_3+\langle\sin(\omega ru)\rangle_4+\langle\sin(\omega ru)\rangle_5\Big)+\langle\cos(\omega ru)\rangle_2\langle\cos(\omega ru)\rangle_5
\\&+\langle\cos(\omega ru)\rangle_3\langle\cos(\omega ru)\rangle_4+\langle\sin(\omega ru)\rangle_2\langle\sin(\omega ru)\rangle_5+\langle\sin(\omega ru)\rangle_3\langle\sin(\omega ru)\rangle_4\bigg\}\,,
\end{split}
\end{align}
\end{widetext}
\end{subequations}
where we have used the Kepler's third law \eqref{Kepler_3rd}, and the angular brackets with subscript `n' represent the integrals as follows:
\begin{subequations}
\begin{align}
\big\langle\cos({\omega r}{u})\big\rangle_{n}\!\equiv\!\int^{\infty}_0\!\!\!\cos\!\bigg(\!\omega r\!\sqrt{1\!+\!\Big(\frac{z}{m_{s}r}\Big)^2}\bigg)\frac{J_1(z)dz}{\big(1\!+\!(\frac{z}{m_{s}r})^2\big)^{\f n2}}\,,
\end{align}
\begin{align}
\big\langle\sin({\omega r}{u})\big\rangle_{n}\!\equiv\!\int^{\infty}_0\!\!\!\sin\!\bigg(\!\omega r\!\sqrt{1\!+\!\Big(\frac{z}{m_{s}r}\Big)^2}\bigg)\frac{J_1(z)dz}{\big(1\!+\!(\frac{z}{m_{s}r})^2\big)^{\f n2}}\,.
\end{align}
\end{subequations}
In order to obtain the total power of scalar radiation we must perform these integrals in the limit $r\rightarrow\infty$. The detailed calculations for these integrals were discussed in Ref. \cite{Alsing2012p64041}. We briefly summarize these calculations in Appendix \ref{appendix_c}, and the results are listed as follows:
\begin{subequations}
\begin{align}
\begin{split}
&\lim\limits_{r\rightarrow\infty}\big\langle\cos({\omega r}{u})\big\rangle_{n}{=}
\\[0.5 em]& \left\{
\begin{matrix}
\cos(\omega r)-v_{s_g}^{n-1}\!(\om)\cos(\omega rv_{s_g}\!(\om))~~~~~~{\rm for}~ \omega>m_s 
\\[0.8 em]
\cos(\omega r)\!-\!\f{(-\!1)^{n\!-\!1}\!+\!1}{2}v_{s_g}^{n\!-\!1}\!(\om)e^{-i\omega rv_{\!s_{\!g}}\!(\om)}~~{\rm for}~ \omega<m_s
\end{matrix}
\right.,
\end{split}
\end{align}
\begin{align}
\begin{split}
&\lim\limits_{r\rightarrow\infty}\big\langle\sin({\omega r}{u})\big\rangle_{n}{=}
\\[0.5 em]& \left\{
\begin{matrix}
\sin(\omega r)-v_{s_g}^{n-1}\!(\om)\sin(\omega rv_{s_g}\!(\om))~~~~~~{\rm for}~ \omega>m_s
\\[0.8 em]
\sin(\omega r)\!-\!\f{(-\!1)^{n\!-\!1}\!-\!1}{2}v_{s_g}^{n\!-\!1}\!(\om)e^{-i\omega rv_{\!s_{\!g}}\!(\om)}~~{\rm for}~ \omega<m_s
\end{matrix}
\right.,
\end{split}
\end{align}
\end{subequations}
where $v_{s_g}\!(\om)=\sqrt{1-m_s^2/\omega^2}$ is the propagation (group) speed of the massive scalar mode (see Eq. \eqref{Scalar_group_phase}).

Performing these integrals in Eqs. \eqref{D_Q_O_scalar_flux}, we obtain the scalar dipole flux
\begin{subequations}\label{D_Q_O_scalar_flux2}
\begin{align}
\begin{split}\label{Dipole_scalar_flux_2}
{\mathcal F}_{\phi}^{D}=&\frac{G(\mathcal Gm\mu)^2}{6R^4}\Big(\E_d^2-\f{G\mu}{R}{\E_d}{\bar{\E}_d}\Big)v_{s_g}^3(\omega)\Theta(\omega-m_s)\,,
\end{split}
\end{align}
the scalar quadrupole flux
\begin{align}
\begin{split}\label{Quadrupole_scalar_flux_2}
{\mathcal F}_{\phi}^{Q}=&\frac{8G\mu^2(\mathcal Gm)^3}{15R^5}\E_q^2v_{s_g}^5(2\omega)\Theta(2\omega-m_s)\,,
\end{split}
\end{align}
and the scalar dipole-octupole flux
\begin{align}
\begin{split}\label{DO_scalar_flux_2}
{\mathcal F}_{\phi}^{DO}=&-\frac{G\mu^2(\mathcal Gm)^3}{30R^5}\E_d\E_ov_{s_g}^5(\omega)\Theta(\omega-m_s)\,,
\end{split}
\end{align}
\end{subequations}
where $\Theta$ is the Heaviside function. Since the screened parameter of the object usually decreases as its mass (or surface gravitational potential) increases, the second term in Eq. \eqref{Dipole_scalar_flux_2} and the dipole-octupole cross term in Eq. \eqref{DO_scalar_flux_2} are the negative modifications to the energy flux at the same PN order as the quadrupole radiation contribution. By comparing Eq. \eqref{Dipole_scalar_flux_2} to Eqs. \eqref{Quadrupole_scalar_flux_2} and \eqref{DO_scalar_flux_2}, we then find that the frequency of the quadrupole scalar wave is twice the frequency of the dipole (or dipole-octupole) scalar wave, which is equal to the orbital frequency in the quasi-circular orbit case. By summing the tensor and scalar energy fluxes \eqref{quadrupole_radiation_flux} and \eqref{D_Q_O_scalar_flux2}, we obtain the total energy fluxes
\begin{align}
\begin{split}
\mathcal F=&{\mathcal F}_g^Q+{\mathcal F}_{\phi}^Q+{\mathcal F}_{\phi}^D+{\mathcal F}_{\phi}^{DO}
\\=&\frac{32G\mu^2(\mathcal Gm)^3}{5R^5}\bigg[1+\frac{1}{12}\E_q^2v_{s_g}^5(2\om)\Theta(2\om-m_{s})
\\&-\frac{1}{192}\E_d\E_ov_{s_g}^5(\om)\Theta(\om-m_{s})
\\&+\frac{5}{192}\left(\frac{R}{Gm}\E_d-\f{\mu}{m}{\bar{\E}_d}\right){\E_d}v_{s_g}^3(\omega)\Theta(\omega-m_{s})\bigg]\,.
\end{split}
\end{align}

Using this and the relations \eqref{binding_energy} and \eqref{E_F}, from Eq. \eqref{P_E} we finally obtain the orbital period decay rate due to the emission of tensor and scalar GWs,
\begin{align}\label{P_P}
\begin{split}
\frac{\dot{P}}{P}=&-\frac{96G\mu(\mathcal Gm)^2}{5R^4}\bigg[1+\frac{1}{12}\E_q^2v_{s_g}^5(2\om)\Theta(2\om-m_{s})
\\&-\frac{1}{192}\E_d\E_ov_{s_g}^5(\om)\Theta(\om-m_{s})
\\&+\frac{5}{192}\left(\frac{R}{Gm}\E_d-\f{\mu}{m}{\bar{\E}_d}\right){\E_d}v_{s_g}^3(\omega)\Theta(\omega-m_{s})
\bigg]\,.
 \end{split}
\end{align}
These results show that in SMG the scalar GWs can be emitted (i.e., scalar mode is excited) if and only if the frequency (energy) of scalar mode is greater than its mass. We know that the Compton wavelength $m_s^{-1}$ is roughly cosmological scales (if $m_s^{-1}\sim1{\rm Mpc}$, then $m_s\sim10^{-14}{\rm Hz}$), and the orbital frequency $\om$ for compact binaries with a 1-hour orbital period is of the order of $10^{-3} {\rm Hz}$, so $m_s\ll\om$ for compact binaries. In this case, the expression \eqref{P_P} for the fractional period derivative can be further simplified to
\begin{align}\label{P_P2}
\begin{split}
\frac{\dot{P}}{P}=&-\frac{96G\mu(\mathcal Gm)^2}{5R^4}\bigg[1+\frac{1}{12}\E_q^2-\frac{1}{192}\E_d\E_o
\\&+\frac{5}{192}\left(\frac{R}{Gm}\E_d-\f{\mu}{m}{\bar{\E}_d}\right){\E_d}\bigg]\,.
\end{split}
\end{align}
Using the Kepler's third law \eqref{Kepler_3rd}, this expression \eqref{P_P2} can be rewritten as
\begin{align}\label{P_P3}
\begin{split}
\dot{P}=-\f{192\pi}{5}\left(\f{2\pi Gm}{P}\right)^{5/3}\!\!\left(\f{\mu}{m}\right){\mathcal A}\,\,,
\end{split}
\end{align}
and we have defined
\begin{align}\label{A_ep}
\begin{split}
\mathcal A=1&+\f{1}{3}{\ep_1\ep_2}+\frac{1}{12}{\E_q^2}-\frac{1}{192}{\E_d\E_o}-\frac{5}{192}\f{\mu}{m}{\E_d}{\bar{\E}_d}
\\&+\frac{5}{192}\Big(\frac{P}{2\pi Gm}\Big)^{2/3}\E_d^2\,.
\end{split}
\end{align}
In Eq. \eqref{A_ep}, the first and second terms represent the contribution of the tensor quadrupole radiation, the third term corresponds to the scalar quadrupole radiation, the fourth term is the contribution of the scalar dipole-octupole cross term, and the last two terms represent the scalar dipole radiation. Because of $Gm/P={\mathcal O}(10^{-9})$ for a typical NS binary with a 1-hour orbital period, the scalar dipole radiation dominates the orbital decay rate, unless $\ep_2-\ep_1\simeq0$\,. In the limiting case ($\ep_1$ and $\ep_2\rightarrow 0$), the expression \eqref{P_P3} reduces to the GR result ($\mathcal A=1$).

\section{Experimental tests in the binary pulsar}\label{section5}
In this section, we discuss how to place constraints on SMG with the orbital decay rate observations of compact binaries. In particular, as an application of our results, we will focus on three specific models of SMG (chameleon, symmetron, and dilaton), and derive the constraints on the model parameters, respectively.

\subsection{Pulsar Constraints}
Up to now, all observations of compact binary systems agree with the GR prediction within observational uncertainties \cite{Stairs2003p5,Wex2014,Manchester2015p1530018,Kramer2016p1630029,Antoniadis2013p6131,Freire2012p3328}. Therefore, in order to place constraints on these gravity theories by using the observations of compact binary systems, the non-GR effects of the theories should be smaller than observational uncertainties.

As mentioned in the previous section, in SMG the scalar dipole radiation dominates the orbital decay rate, unless $\ep_2-\ep_1\simeq0$. Due to the large difference of the screened parameters in the neutron star-white dwarf (NS-WD) binary systems, these systems are the best target to constrain the model parameters in SMG. Now, let us consider a NS-WD binary system with quasi-circular orbit. The screened parameter is inversely proportional to the surface gravitational potential (see Eq. \eqref{epsilon_a}), i.e., $\ep_{\rm WD}/\ep_{\rm NS}\simeq\Phi_{\rm NS}/\Phi_{\rm WD}\sim10^{4}$\,. Therefore, the difference in the screened parameters is approximately equal to the WD screened parameter, i.e., $\E_d=\ep_{\rm WD}-\ep_{\rm NS}\simeq\ep_{\rm WD}$\,. Using these, the expression \eqref{A_ep} can be simplified to
\begin{align}\label{A_ep_NSWD}
\begin{split}
\mathcal A=1+\frac{5}{192}\Big(\frac{P}{2\pi Gm}\Big)^{2/3}\ep^{2}_{\rm WD}\,.
\end{split}
\end{align}
We can also write the observed value $\mathcal A^{\rm obs}$  as
\begin{align}\label{A_ep_NSWD_obs}
\begin{split}
\mathcal A^{\rm obs}=\f{\dot{P}^{\rm obs}}{\dot{P}^{\rm GR}}=1+\delta\pm\sigma\,,
\end{split}
\end{align}
where $\delta$ is the fractional deviation of the observed value from the GR prediction, and $\sigma$ is the observational uncertainty. Comparing Eq. \eqref{A_ep_NSWD_obs} with Eq. \eqref{A_ep_NSWD}, we obtain the constraint
\begin{align}\label{A_ep_NSWD_obs4}
\begin{split}
\left|\frac{5}{192}\Big(\frac{P}{2\pi Gm}\Big)^{2/3}\ep^{2}_{\rm WD}-\delta\right|\le2\sigma\,
\end{split}
\end{align}
at 95\% confidence level (CL). This constraint relation can be further simplified to
\begin{align}\label{ep_WD_obs1}
\begin{split}
\ep_{\rm WD}\le(\delta+2\sigma)^{1/2}\left(\f{m}{P}\right)^{1/3}\times1.269\times10^{-2}\,
\end{split}
\end{align}
at 95\% CL, where the total mass $m$ is expressed in units of solar masses, and the orbital period $P$ is expressed in units of hours. For the general SMG, including chameleon, symmetron, and dilaton theories, $\phi_{\rm min}(\rho)$ (in Eq.\eqref{mass_eff}) is generally inversely correlated to  the matter density $\rho$ of the local environment of the scalar field. The background matter density is always much less than the WD density, i.e., $\rho_{b}\ll\rho_{\rm WD}$, so we have $\phi_{\rm VEV}\gg\phi_{\rm WD}$ \cite{Zhang2016p124003}. Using this and substituting the relation \eqref{epsilon_a} into the constraint \eqref{ep_WD_obs1}, we obtain 
\begin{align}\label{ep_WD_obs5}
\begin{split}
\f{\phi_{\rm VEV}}{\Mpl}\le(\delta+2\sigma)^{1/2}\left(\f{m}{P}\right)^{1/3}\f{m_{\rm WD}}{R_{\rm WD}}\times2.694\times10^{-8}\,
\end{split}
\end{align}
at 95\% CL, where the mass $m_{\rm WD}$ and radius $R_{\rm WD}$ of the WD are expressed in units of solar masses and solar radii, respectively.

\begin{center}
\begin{table}[htb]
\caption{Parameters relevant to the binary system PSR J1738+0333 \cite{Freire2012p3328}.}
\label{PSR J1738+0333_parameters}
\begin{tabular}{l r}
\hline
\hline
Eccentricity, $e$  & $(3.4\pm1.1)\times10^{-7}$ \\
Period, $P$ (day) & 0.3547907398724(13) \\
Period derivative, $\dot{P}^{\rm obs}$ & $(-25.9\pm3.2)\times10^{-15}$ \\
$\dot{P}^{\rm obs}/\dot{P}^{\rm GR} $ & $0.93\pm0.13$\\
Total Mass, $m$ ($M_\odot$) & $1.65^{+0.07}_{-0.06}$ \\
WD Mass, $m_{\rm WD}$ ($M_\odot$) & $0.181^{+0.008}_{-0.007}$ \\
White dwarf radius, $R_{\rm WD}$ ($R_\odot$) & $0.037^{+0.004}_{-0.003}$\\
\hline
\hline 
\end{tabular}
\end{table}
\end{center}

In this paper, we use the observation data of the binary system PSR J1738+0333 which is a 5.85-ms pulsar in a 8.51-hour quasi-circular orbit with a low-mass WD companion \cite{Freire2012p3328,Antoniadis2012p3316}. 
The orbital parameters for this system are listed in Table \ref{PSR J1738+0333_parameters}, which are taken directly from \cite{Freire2012p3328}. Using these observed values of the orbital parameters, from the constraints \eqref{ep_WD_obs1} and \eqref{ep_WD_obs5} we obtain an upper bound on the WD screened parameter
\begin{align}\label{ep_WD_obs}
\begin{split}
\ep_{\rm WD}\le3.2\times10^{-3}\,
\end{split}
\end{align}
at 95\% CL, and an upper bound on the scalar field VEV
\begin{align}\label{phi_VEV_obs}
\begin{split}
\f{\phi_{\rm VEV}}{\Mpl}\le3.3\times10^{-8}\,
\end{split}
\end{align}
at 95\% CL.
%As one of the main conclusions of this article, these results are quite general, which are applicable for any SMG models. 
In the following subsections, we shall apply them to three specific models of SMG (chameleon, symmetron, and dilaton), and derive the constraints on the model parameters by the pulsar observations. For comparison, we will also present the constraints on these three models by the observations in solar system \cite{Bertotti2003p374}.

\subsection{Chameleon}
The chameleon model was introduced as a screening mechanism by Khoury and Weltman \cite{Khoury2004p171104,Khoury2004p44026,Gubser2004p104001}. The chameleon mechanism operates a thin-shell shielding scalar field, which acquires a large mass in dense environments and suppresses its ability to mediate a fifth force. The original chameleon is ruled out by the combined constraints of the solar system and cosmology \cite{Hees2012p103005, Zhang2016p124003}. Here, we consider the exponential chameleon, which is characterized by an exponential potential and an exponential coupling function \cite{modified_chamelons},
\begin{subequations}\label{Chameleon_V_A}
\begin{align}
V(\phi)=\Lambda^4\exp\Big(\frac{\Lambda^{\alpha}}{\phi^{\alpha}}\Big)\,, 
\end{align}
\begin{align}
A(\phi)=\exp\Big(\frac{\beta\phi}{M_\text{Pl}}\Big)\,, 
\end{align}
\end{subequations}
where $\beta$ is a positive dimensionless coupling constant, $\alpha$ is a positive dimensionless constant index, and $\Lambda$ labels the energy scale of the theory and today is close to the dark energy scale ($\Lambda=2.24\times10^{-3}$\,eV) \cite{Hamilton2015p849,Zhang2016p124003}.

\begin{figure}
\begin{center}
\includegraphics[width=8cm, height=4.5cm]{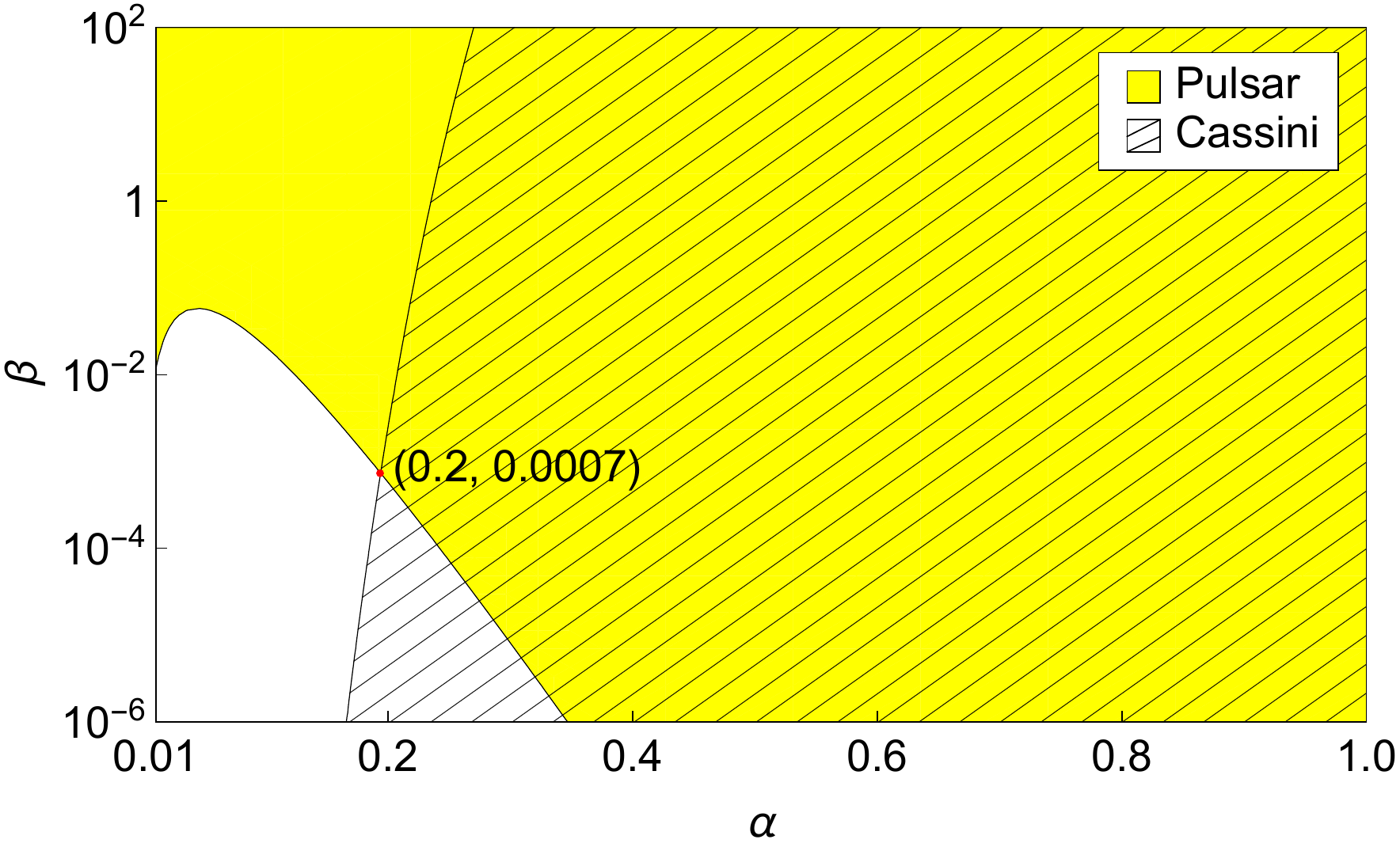}
\caption{In the parameter space of exponential chameleon model, the shadow region is allowed by the tests of Cassini experiment, while the yellow region is allowed by the observations of PSR J1738+0333\,. The combined constraints of the two experiments require $\alpha\ge0.2$\,.}\label{chameleon_alpha_beta}
\end{center}
\end{figure}

Substituting chameleon potential and coupling function \eqref{Chameleon_V_A} into Eq. \eqref{Veff2}, from Eq. \eqref{mass_eff} we have the chameleon VEV and mass, 
\begin{subequations}\label{chameleon_VEV_mass}
\begin{align}
\phi_{\rm VEV}=\bigg(\frac{{\alpha} M_\text{Pl} \Lambda^{4+\alpha}}{\beta\rho_{b}}\bigg)^{\frac{1}{\alpha+1}}\label{chameleon_VEV}\,,
\end{align}
\begin{align}
m^2_{s}=\frac{(\alpha+1)\beta\rho_{b}}{M_\text{Pl}\phi_{\rm VEV}}+\f{\beta^2\rho_{b}}{\Mpl^2}\label{chameleon_mass}\,.
\end{align}
\end{subequations}
Here, $\rho_{b}$ is the background matter density, and $\rho_{b}=\rho_{gal}\simeq10^{-42}\,\rm GeV^4$ which roughly corresponds to the galactic matter density. Using the pulsar constraint \eqref{phi_VEV_obs}, from Eq. \eqref{chameleon_VEV} we derive the following relation between $\alpha$ and $\beta$,
\begin{align}\label{chameleon_beta_alpha}
\log\beta\ge\log\alpha-22.6\alpha+2.88\,,
\end{align}
which is illustrated in Fig. \ref{chameleon_alpha_beta} by the yellow region. In addition, for the chameleon model, the PPN parameter $\gamma=1-2\beta\ph/(M_\text{Pl}\Phi)$ (see \cite{Zhang2016p124003} for detailed derivations), from the Cassini constraint $\left|\gamma_{\rm obs}-1\right|\le2.3\times10^{-5}$ \cite{Bertotti2003p374}, we present the allowed region in the parameter space $(\alpha,\,\beta)$ in Fig. \ref{chameleon_alpha_beta} by the shadow region.

Fig. \ref{chameleon_alpha_beta} shows the bound on the model parameters $\alpha$ and $\beta$ by considering the galactic background. The yellow region is allowed by the orbital decay rate observations of PSR J1738+0333. The shadow region indicates the parameter space allowed by Cassini experiment in the solar system. The overlap region allowed by the combined constraints of the two experiments gives the stringent bound $\alpha\ge0.2$\,.

\begin{figure}
\begin{center}
\includegraphics[width=8cm, height=4.5cm]{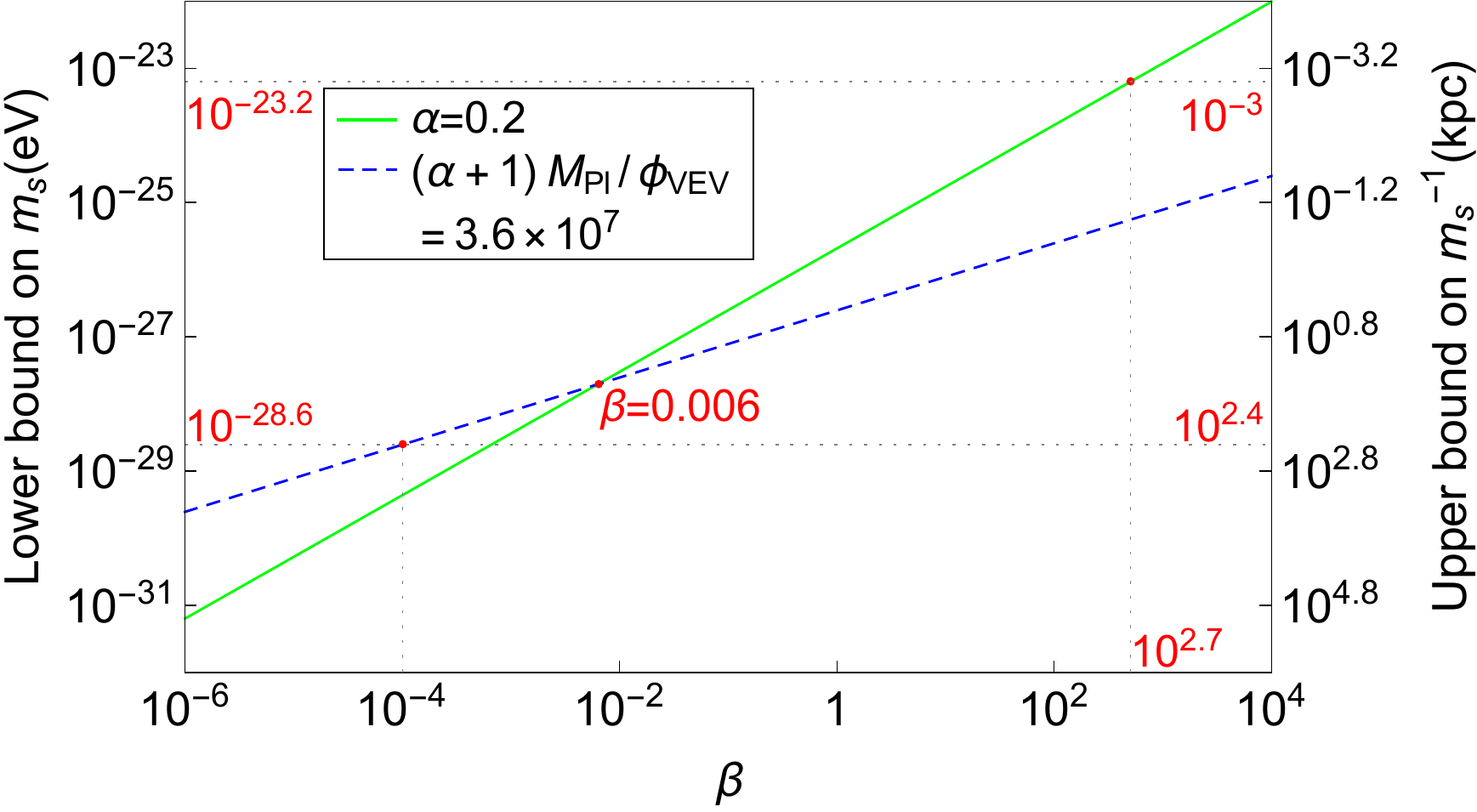}
\caption{Lower bound on chameleon mass $m_s$ (upper bound on $m_s^{-1}$) as a function of the coupling constant $\beta$ from the constraints $\alpha\ge0.2$ (green solid line) and Eq. \eqref{chameleon_alpha_phi} (blue dashed line)\,.}\label{chameleon_ms_beta}
\end{center}
\end{figure}

Substituting the chameleon VEV  \eqref{chameleon_VEV}  into the chameleon mass \eqref{chameleon_mass} and imposing the constraint $\alpha\ge0.2$ yields the lower bound on chameleon mass $m_s$ in Fig. \ref{chameleon_ms_beta} by the green solid line. By combining $\alpha\ge0.2$ and the pulsar constraint \eqref{phi_VEV_obs} yields the constraint 
\begin{align}\label{chameleon_alpha_phi}
\f{(\alpha+1)\Mpl}{\ph}\ge3.6\times10^7\,.
\end{align}
Using this, from Eq. \eqref{chameleon_mass} we obtain the lower bound on chameleon mass $m_s$ in Fig. \ref{chameleon_ms_beta} by the blue dashed line. 

Fig. \ref{chameleon_ms_beta} shows the lower bound on chameleon mass $m_s$ as a function of the coupling constant $\beta$\,. The green solid and blue dashed lines indicate the bounds on $m_s$ from the constraints $\alpha\ge0.2$ and Eq.  \eqref{chameleon_alpha_phi}\,, respectively. From Fig. \ref{chameleon_ms_beta} we can see that $m_s\ge10^{-28.6} {\rm~ eV}$ ($m_s^{-1}\le10^{2.4}{\rm~ kpc}$) if $\beta\ge10^{-4}$, and $\beta\le10^{2.7}$ if $m_s^{-1}\ge1{\rm~ pc}$\,.

\subsection{Symmetron}
The symmetron models are characterized by a $\mathbb Z_2$ symmetry breaking potential (a mexican hat potential) and a quadratic coupling function \cite{Hinterbichler2010p231301,Hinterbichler2011p103521,Davis2012p61},
\begin{subequations}\label{Symmetron_V_A}
\begin{align}
V(\phi)=V_0-\frac12\mu^2\phi^2+\frac{\lambda}{4}\phi^4\,,
\end{align}
\begin{align}
A(\phi)=1+\frac{\phi^2}{2M^2}\,,
\end{align}
\end{subequations}
where $\mu$ and $M$ are mass scales,  $\lambda$ is a positive dimensionless coupling constant, $V_0$ is the vacuum energy of the bare potential $V(\phi)$.  In high density regions the $\mathbb Z_2$ symmetry is unbroken and the fifth force is absent, whereas in low density regions the $\mathbb Z_2$ symmetry is spontaneously broken and the fifth force is present.

\begin{figure}
\centering
\subfigure{
\begin{minipage}[b]{0.45\textwidth}
\includegraphics[width=1\textwidth, height=0.43\textwidth]{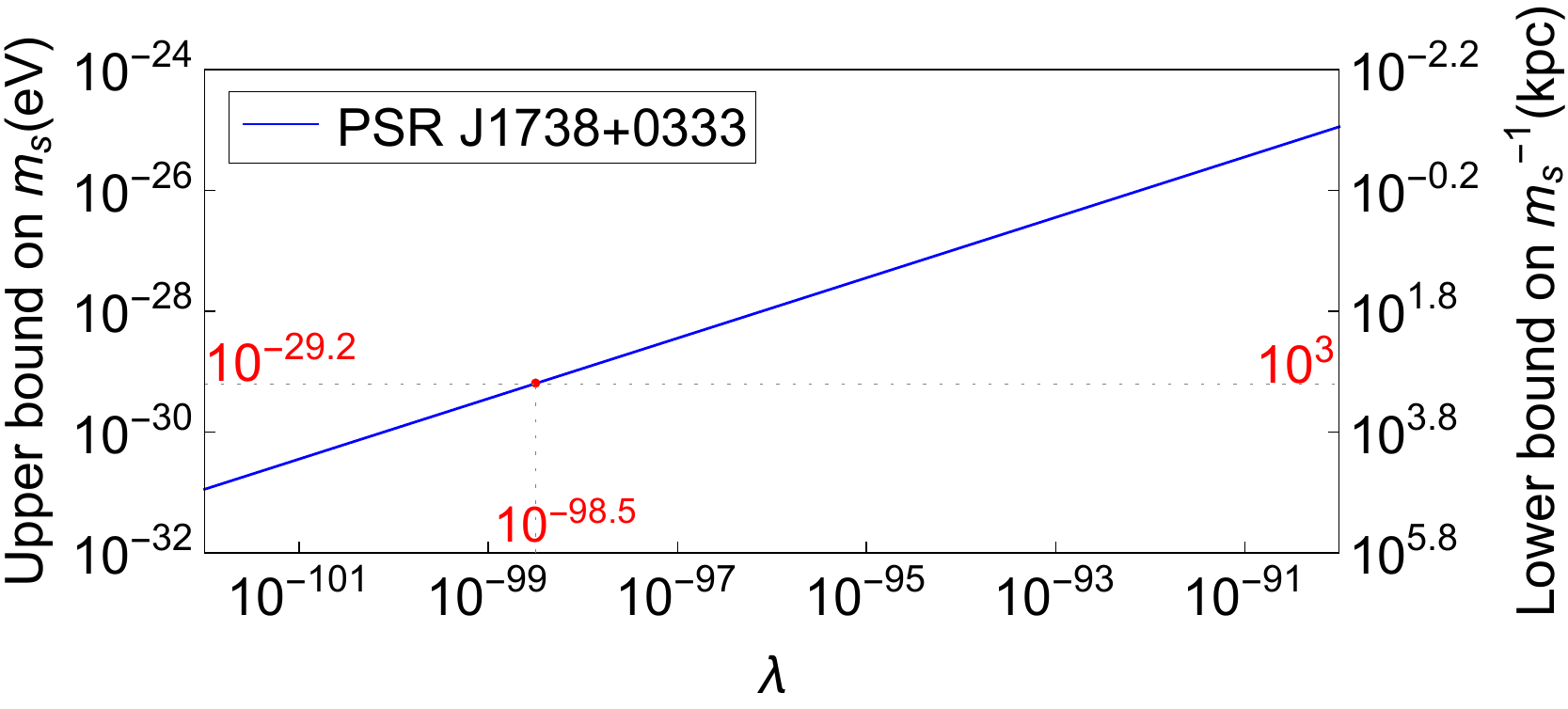}
\end{minipage}
}
\subfigure{
\begin{minipage}[b]{0.45\textwidth}
\includegraphics[width=1\textwidth, height=0.43\textwidth]{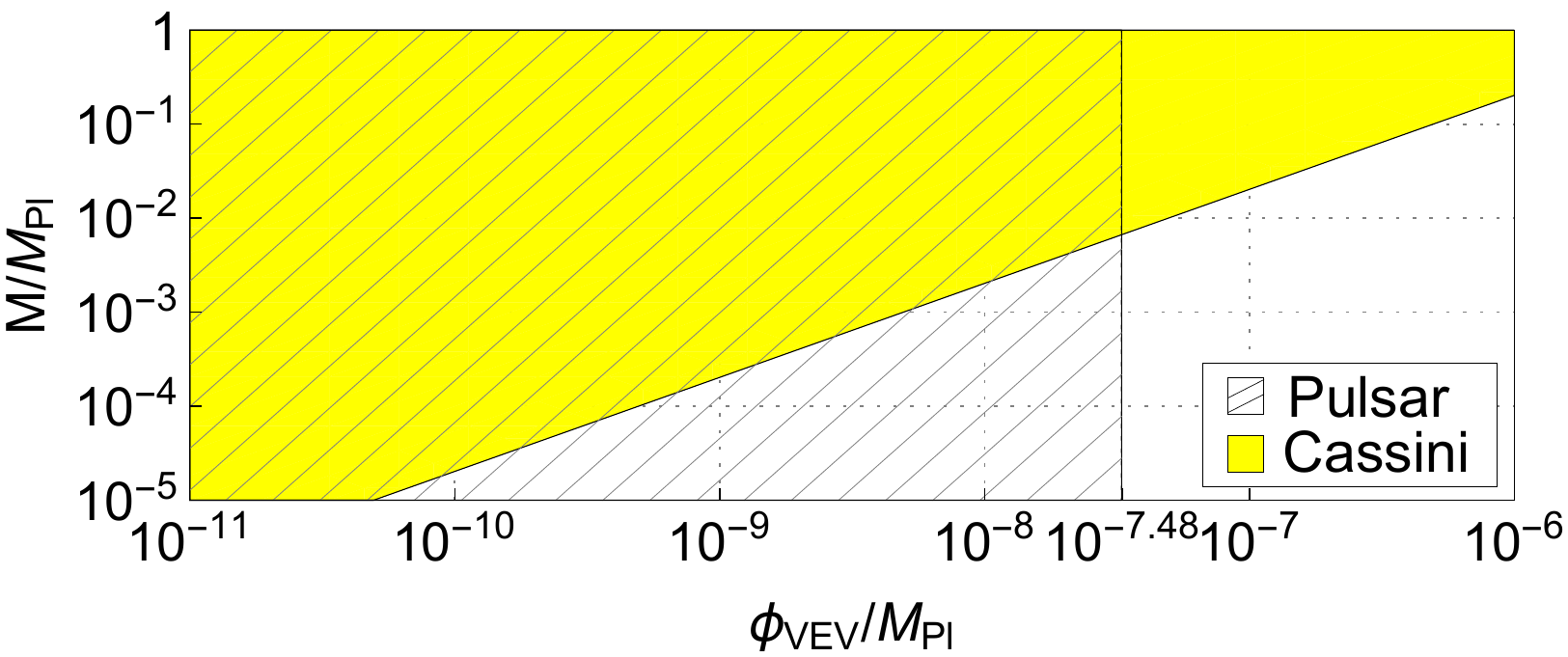}
\end{minipage}
}
 \caption{Top: Upper bound on symmetron mass $m_s$ (lower bound on $m_s^{-1}$) as a function of the coupling constant $\lambda$ from the observations of PSR J1738+0333\,.
\\Bottom: In the parameter space of symmetron model, the yellow region is allowed by the tests of Cassini experiment, while the shadow region is allowed by the observations of PSR J1738+0333\,. 
} \label{symmetron_fig}
\end{figure}

Substituting symmetron potential and coupling function \eqref{Symmetron_V_A} into Eq. \eqref{Veff2}, from Eq. \eqref{mass_eff} we obtain the relation between the symmetron VEV $\ph$ and the symmetron mass $m_s$, 
\begin{align}\label{symmetron_VEV}
\phi_{\rm VEV}=\f{m_s}{\sqrt{2\lambda}}\,.
\end{align}
Using this, from the pulsar constraint \eqref{phi_VEV_obs} we obtain the upper bound on symmetron mass $m_s$ in the top plot of Fig. \ref{symmetron_fig}\,. For the symmetron model, the PPN parameter $\gamma=1-2\ph^2/(M^2\Phi)$ (see \cite{Zhang2016p124003} for detailed derivations), from the Cassini constraint $\left|\gamma_{\rm obs}-1\right|\le2.3\times10^{-5}$ \cite{Bertotti2003p374} and the pulsar constraint \eqref{phi_VEV_obs}, we obtain the combined constraints on the parameter space ($\ph,\,M$)\,. This result is displayed in the bottom plot of Fig. \ref{symmetron_fig}\,.

The top panel of Fig. \ref{symmetron_fig} shows the upper bound on symmetron mass $m_s$ as a function of the coupling constant $\lambda$ of $\phi^4$ interaction, which is derived from the orbital decay rate observations of PSR J1738+0333. From the top panel we find a relatively weak bound $\lambda\ge10^{-98.5}$, if $m_s^{-1}\le10^{3}\,\, {\rm kpc}$\,. The bottom panel shows the bound on the parameter space ($\ph,\,M$)\,. The yellow region is allowed by the tests of Cassini experiment in the solar system, while the shadow region indicates the pulsar constraint \eqref{phi_VEV_obs} from the orbital decay rate observations of PSR J1738+0333.

\subsection{Dilaton}
The dilaton model, inspired by string theory in the large string coupling limit, has an exponentially runaway potential and a quadratic coupling function \cite{Damour1994p532,Damour1994p1171,Brax2010p63519},
\begin{subequations}\label{Dilaton_V_A}
\begin{align}
V(\phi)= V_0\exp\Big(-\frac{\phi}{M_\text{Pl}}\Big)\,,
\end{align}
\begin{align}
A(\phi)=1+\frac{(\phi-\phi_\star)^2}{2M^2}\,,
\end{align}
\end{subequations}
where $V_0$ is a constant with the dimension of energy density, $M$ labels the energy scale of the theory, and $\phi_\star$ is approximately the value of $\phi$ today. The dilaton mechanism is similar to the symmetron. The coupling between dilaton and matter is negligible in dense regions, while in low density regions the dilaton mediates a gravitational-strength fifth force.

Substituting dilaton potential and coupling function \eqref{Dilaton_V_A} into Eq. \eqref{Veff2}, from Eq. \eqref{mass_eff} we have the dilaton VEV and mass, 
\begin{subequations}\label{dilaton_VEV_mass}
\begin{align}
\phi_{\rm VEV}=\phi_\star+\frac{M^2\rho_{\Lambda_0}}{M_\text{Pl}\rho_{b}}\label{dilaton_VEV}\,,
\end{align}
\begin{align}
m^2_{ s}=\frac{\rho_{b}}{M^2}+\f{\rho_{\Lambda_0}}{\Mpl^2}\label{dilaton_mass}\,.
\end{align}
\end{subequations}
Using the pulsar constraint \eqref{phi_VEV_obs}, from Eq. \eqref{dilaton_VEV_mass} we derive the constraint on model parameters,
\begin{subequations}\label{dilaton_M_mass_pulsar}
\begin{align}
\f{M}{\Mpl}\le0.036
\end{align}
and
\begin{align}
m_{ s}\ge1.1\times10^{-29}\,{\rm eV}\quad({\rm or}\quad m^{-1}_{ s}\le0.58\,\rm Mpc)
\end{align}
\end{subequations}
at 95\% CL.
For the dilaton model, the PPN parameter $\gamma=1-2(\ph-\phi_\star)^2/(M^2\Phi)$ (see \cite{Zhang2016p124003} for detailed derivations), from the Cassini constraint $\left|\gamma_{\rm obs}-1\right|\le2.3\times10^{-5}$ \cite{Bertotti2003p374} we have 
\begin{subequations}\label{dilaton_M_mass_Cassini}
\begin{align}
\f{M}{\Mpl}\le0.20
\end{align}
and
\begin{align}
m_{ s}\ge2.1\times10^{-30}\,{\rm eV}\quad ({\rm or}\quad m^{-1}_{ s}\le3.1\,\rm Mpc)
\end{align}
\end{subequations}
at 68\% CL.
By comparing Eq. \eqref{dilaton_M_mass_pulsar} with Eq. \eqref{dilaton_M_mass_Cassini}, we find that the pulsar constraints are more stringent than the solar system tests.
%we find that the pulsar constraints are not competitive with the solar system tests.

\section{Conclusions}\label{section6}

The salient feature of SMG is the screening mechanism, which can suppress the fifth force in dense regions and allow theories to evade the tight gravitational tests in the solar system and the laboratory. In this paper, we investigated how the screening mechanisms in SMG affect the orbital evolution of compact binaries due to the tensor and scalar gravitational radiations, and derived the constraints on the screening mechanisms by the pulsar observations. In any theory of gravity, the gravitational radiation generally depends not only on the dissipative sector which regulates how fast the system loses energy, but also on the conservative sector which regulates the orbital dynamics of the system. In alternative theories (including SMG), both the sectors are generally modified by the additional fields controlled by the sensitivities, which characterize how the gravitational binding energy of a compact object responds to its motion relative to the additional fields.

In SMG, we first considered the modifications to the conservative sector of the theory. By solving the PN equations for the massless tensor and massive scalar fields in the near zone, we derived the EIH equations of motion for a compact binary system. It turned out that both the WEP and the gravitational inverse-square law are violated in general. However, in the near zone, the inverse-square law can be approximately satisfied, which guarantees that the Kepler's third law holds. In addition, by comparing with the two scalar solutions obtained by means of different methods, we found that the first sensitivity is completely equivalent to the screened parameter.

In the dissipative sector, we solved the wave equations for the massless tensor and massive scalar fields in the wave zone, calculated in detail the rate of the energy loss due to the emission of tensor and scalar GWs, and derived their contributions to the change in the orbital period. The tensor radiation in SMG behaves as in GR at leading PN order, and there is neither monopole nor dipole radiations. The emission of scalar radiation starts at monopole order, but there is no monopole contribution to leading order in the quasi-circular orbit case. The scalar dipole radiation depends not only on the difference in screened parameters but also on the propagation speed of the massive scalar particle. The dipole-octupole cross term appearing in the scalar radiation is the negative modification to the energy flux at the same PN order as the quadrupole radiation contribution. We focused mostly on the scalar dipole radiation, which is generally stronger than quadrupole radiation and leads to a strong modification to the evolution of the orbital period.

In SMG, all modifications (of the conservative and dissipative sectors) are due to the scalar field controlled by the object's screened parameter (or scalar charge), which is inversely proportional to the object's surface gravitational potential. For the compact objects (such as white dwarfs and neutron stars), the effects of the scalar sector of SMG are strongly suppressed by the screening mechanisms, and thus the deviations from GR become small and weak. In other words, SMG looks more like GR for strongly self-gravitating bodies, which is completely different from other alternative theories without screening mechanisms.

All current pulsar observations agree with GR's predictions within the observational uncertainties \cite{Stairs2003p5,Wex2014,Manchester2015p1530018,Kramer2016p1630029,Antoniadis2013p6131,Freire2012p3328}, which allows us to place the stringent constraints on the screening mechanisms in SMG. By comparing our results for the orbital period decay rate to the observations of quasi-circular binary system PSR J1738+0333, we obtained the quite stringent bounds on the screened parameter and the scalar field VEV.

Finally, we applied our results to three specific models of SMG (chameleon, symmetron, and dilaton), and derived the pulsar constraints on the model parameters, respectively. For comparison, we also discussed the solar system constraints on these three models. Consistent with all the previous works, we found the following results for these SMG models: The combined observations of the pulsar and solar systems yield a lower bound on the chameleon parameter $\alpha$ and a lower bound on the chameleon mass $m_s$ as a function of the chameleon coupling constant $\beta$. Contrary to chameleon, the pulsar observations yield an upper bound on the symmetron mass $m_s$ as a function of the symmetron coupling constant $\lambda$. For the dilaton model, the pulsar constraints are more stringent than the solar system tests. All these models pass the current constraints from the pulsar and solar systems, and we obtained the bounds on the model parameters, respectively.

At the end of this paper, we would like to emphasize that the results derived in this article are applicable for the quasi-circular orbits of compact binary system. In a separate paper, we will extend these calculations to much more general case with the quasi-elliptic orbits.

\begin{acknowledgments}
We appreciate the helpful discussion with Anzhong Wang, Liming Cao, Kejia Lee, Yuxiao Liu, Xian Gao, He Huang, and Yifu Cai. This work is supported by NSFC No. 11603020, 11633001, 11173021, 11322324, 11653002, 11421303, project of Knowledge Innovation Program of Chinese Academy of Science, the Fundamental Research Funds for the Central Universities and the Strategic Priority Research Program of the Chinese Academy of Sciences Grant No. XDB23010200.
\end{acknowledgments}

\appendix

\section{Scalar solution}\label{appendix_a}
Now let us solve the scalar field equation by the method of matching the internal and external solutions. We consider a static spherically symmetric source object with constant density $\rho_o$ and radius $R$, which is embedded in a homogeneous background of matter density $\rho_b$. Then, the scalar field equation \eqref{scalar_eom} can be simplified to
\be
\label{scalar field eq.3}
\frac{{\rm d}^2\phi}{{\rm d}r^2}+\frac 2r\frac{{\rm d}\phi}{{\rm d}r}=m^2_{\rm m}(\rho)\big[\phi-\phi_{\rm m}(\rho)\big]\,,
\ee
with
\begin{equation}
\rho(r) = \left\{
\begin{matrix}
\rho_{o}\qquad~{\rm for}~~~ r<R \cr
\rho_{b}\qquad~ {\rm for}~~~ r>R
\end{matrix}
\right..
\end{equation}
%\end{subequations}
This is a second order differential equation, and as such we must impose two boundary conditions. The first is that the solution is regular at the origin, i.e., ${\rm d\phi}/{\rm d}r\big|_{r=0} = 0$, and the second is that the scalar field asymptotically converges to the scalar background, i.e.,  $\phi\big|_{r\rightarrow\infty}\rightarrow\ph$. Moreover, $\phi$ and ${\rm d}\phi/{\rm d}r$ are of course continuous at the surface of the  object. By solving Eq. \eqref{scalar field eq.3} directly, we get the exact solution
\begin{subequations}
\begin{align}
\phi(r<R)&=\phi_o+\frac Ar\sinh(m_or)\,,
\\
\label{exterior scalar field}
\phi(r>R)&=\ph+\frac Bre^{-m_s r}\,,
\end{align}
\end{subequations}
with
\begin{subequations}
\begin{align}
A&=\frac{(\ph-\phi_o)(1+m_s R)}{m_o\cosh(m_oR)+m_s\sinh(m_oR)}\,,
\\
B&=-e^{m_s R}(\ph-\phi_o)\frac{m_oR-\tanh(m_oR)}{m_o+m_s\tanh(m_oR)}\,,
\end{align}
\end{subequations}
where $\phi_o$ and $\ph$ are respectively the positions of the minimum of $V_{\rm eff}$ inside and far outside the source object, $m_o$ and $m_s$ are respectively the effective masses of the scalar field at $\phi_o$ and $\ph$.
In general, the radius $R$ is much larger than the fifth force range $m^{-1}_o$, but is much less than the Compton wavelength $m_s^{-1}$, that is $m_o^{-1}\ll R\ll m_s^{-1}$. Using this, the exterior scalar field \eqref{exterior scalar field} reduces to
%\begin{subequations}
\begin{align}\label{exterior scalar solution}
\begin{split}
\varphi(r)&=\phi(r)-\ph
\\&=- M_\text{Pl}\frac{Gm\epsilon}re^{-m_s r}\,,
\end{split}
\end{align}
with
\be\label{epsilon}
\epsilon\equiv\frac{\ph-\phi_o}{M_\text{Pl}\Phi},
\ee
%\end{subequations}
where $m$ is the mass of the object, $\Phi=Gm/R$ is its surface gravitational potential. The quantity $\epsilon$ is always called the screened parameter (or scalar charge) of the object. 
In the case of $\epsilon\ll1$, the scalar field is strongly suppressed (i.e., the screening mechanism is very strong), whereas in the case of $\epsilon\gtrsim1$, the screening mechanism is weak and the scalar force (fifth force) becomes comparable with the gravitational force.

For a multibody system, we have the scalar field
\begin{align}
\varphi=\sum_a\varphi_a=-\Mpl\sum_a\frac{Gm_a\ep_a}{r_a}e^{-m_sr_a}\label{multibody_varphi}\,,
\end{align}
where $m_a$ is the mass of the $a$-th object, $\ep_a$ is its screened parameter, $m_s$ is the effective mass of the scalar, and $r_a=\left|\mathbf{r}-\mathbf{r}_a(t)\right|$\,.

\section{PN expansion of the metric tensor}\label{appendix_b}

Here we will derive in detail the PN expansion \eqref{PN_metric} of the metric tensor. We follow very closely the method outlined in \cite{Will1993p}. For convenience, the tensor field equations \eqref{tensor_eom} is written in the equivalent form
\begin{align}
R_{\mu\nu}\label{tensor_eom_R}
= 8 \pi G \left[S_{\mu\nu}+\partial_\mu\phi\partial_\nu\phi+V(\phi)g_{\mu\nu}\right]\,,
\end{align}
with 
\begin{align}
S_{\mu\nu}\label{tensor_S_munu}
\equiv T_{\mu\nu}-\f12g_{\mu\nu}T\,,
\end{align}
where $T_{\mu\nu}$ and $T$ are respectively the energy-momentum tensor of matter and its trace, given in Eqs. \eqref{Tuv_matter} and \eqref{Tm_trace}. 

In the weak-field limit around the flat Minkowski background and the scalar field VEV (scalar background), the tensor $S_{\mu\nu}$ is expanded in the form:
\begin{subequations}
\begin{align}
\begin{split}
S_{00} =&\frac{1}{2}\sum_am_a\delta^3(\mathbf{r}-\mathbf{r}_a)\Big(1+\f{3}{2}v^2_a-\accentset{(2)}h_{00}
\\&-\f{1}{2}\accentset{(2)}h_{ij}\delta_{ij}+\f{\epsilon_a}{2\Mpl}\accentset{(2)}\varphi\Big)+\mathcal{O}(v^6)\,,
\end{split}
\\
S_{0j} =&-\sum_am_av_a^j\delta^3(\mathbf{r}-\mathbf{r}_a)
+ \mathcal{O}(v^5)\,,
\\
S_{ij} =&\f{\delta_{ij}}{2}\sum_am_a\delta^3(\mathbf{r}-\mathbf{r}_a)
+ \mathcal{O}(v^4)\,.
\end{align}
\end{subequations}
By using these relations, the right-hand sides of the tensor field equations \eqref{tensor_eom_R} can be expanded to the required order in the form:
\begin{subequations}\label{R_right}
\begin{align}
\begin{split}
R_{00}=&4\pi G\sum_am_a\delta^3(\mathbf{r}-\mathbf{r}_a)\Big(1+\f{3}{2}v^2_a-\accentset{(2)}h_{00}
\\&-\f{1}{2}\accentset{(2)}h_{ij}\delta_{ij}+\f{\epsilon_a}{2\Mpl}\accentset{(2)}\varphi\Big)+\mathcal{O}(v^6)\,,\label{R00_right}
\end{split}
\\
\begin{split}
R_{0j}=&-8\pi G\sum_am_av_a^j\delta^3(\mathbf{r}-\mathbf{r}_a)+\mathcal{O}(v^5)\,,\label{R0j_right}
\end{split}
\\
\begin{split}
R_{ij}=&4\pi G\delta_{ij}\sum_am_a\delta^3(\mathbf{r}-\mathbf{r}_a)+\mathcal{O}(v^4)\,,\label{Rij_right}
\end{split}
\end{align}
\end{subequations}
where we have neglected the bare potential $V(\phi)$ corresponding to the dark energy. 
The left-hand sides of the tensor field equations \eqref{tensor_eom_R}, i.e., the components of the Ricci tensor, are expanded to the same order in the form:
\begin{subequations}\label{R_left}
\begin{align}
\begin{split}
R_{00}=&\!-\!\frac{1}{2}\nabla^2\ac{(2)}h_{00}\!-\!\frac{1}{2}\nabla^2\ac{(4)}h_{00}\!-\!\f{1}{4}\!\big(\!\boldsymbol{\nabla}\ac{(2)}h_{00}\!\big)^2\!\!-\!\frac{1}{2}\!\Big(\ac{(2)}h_{jj,00}\!-\!2\ac{(3)}h_{j0,j0}\!\Big)
\\&+\!\frac{1}{2}\ac{(2)}h_{00,j}\!\Big(\ac{(2)}h_{jk,k}\!-\!\f{1}{2}\ac{(2)}h_{kk,j}\!\Big)\!+\!\f{1}{2}\ac{(2)}h_{jk}\ac{(2)}h_{00,jk}\!+\!\mathcal{O}(v^6)\label{R00_left}\,,
\end{split}
\\
R_{0j}=&\!-\!\frac{1}{2}\!\Big(\!\nabla^2\ac{(3)}h_{0j}\!+\!\ac{(2)}h_{kk,0j}\!-\!\ac{(3)}h_{k0,jk}\!-\!\ac{(2)}h_{kj,0k}\!\Big)\!+\!\mathcal{O}(v^5)\label{R0j_left}\,,
\\
R_{ij}=&\!-\!\frac{1}{2}\!\Big(\!\nabla^2\ac{(2)}h_{ij}\!-\!\ac{(2)}h_{00,ij}\!+\!\ac{(2)}h_{kk,ij}\!-\!\ac{(2)}h_{ki,kj}\!-\!\ac{(2)}h_{kj,ki}\!\Big)\!+\!\mathcal{O}(v^4)\label{Rij_left}\,.
\end{align}
\end{subequations}
In addition, in order to solve the tensor field equations, we generally impose the PN gauge condition \cite{Will1993p}
\begin{subequations}\label{PN_gauge}
\begin{align}
&h^{\mu}_{i,\mu}-\f{1}{2}h^{\mu}_{\mu,i}=0\label{PN_gauge_i}\,,
\\
&h^{\mu}_{0,\mu}-\f{1}{2}h^{\mu}_{\mu,0}=-\f{1}{2}h_{00,0}\label{PN_gauge_0}\,.
\end{align}
\end{subequations}

We consider the PN tensor field equations \eqref{R00_right} and \eqref{R00_left} up to order $\mathcal{O}(v^2)$, and obtain the equation
\begin{align}
\begin{split}
\nabla^2\ac{(2)}h_{00}=-8\pi G\sum_am_a\delta^3(\mathbf{r}-\mathbf{r}_a)\,,
\end{split}
\end{align}
this solution is 
\begin{align}
\begin{split}
\ac{(2)}h_{00}=2\sum_a\f{Gm_a}{r_a}\label{h_00_2}\,.
\end{split}
\end{align}
For the spatial components, up to order $\mathcal{O}(v^2)$, using the PN gauge \eqref{PN_gauge_i}, the PN tensor field equations \eqref{Rij_right} and \eqref{Rij_left} follow
\begin{align}
\begin{split}
\nabla^2\ac{(2)}h_{ij}=-8\pi G\delta_{ij}\sum_am_a\delta^3(\mathbf{r}-\mathbf{r}_a)\,,
\end{split}
\end{align}
this solution is 
\begin{align}
\begin{split}
\ac{(2)}h_{ij}=2\delta_{ij}\sum_a\f{Gm_a}{r_a}\label{h_ij_2}\,.
\end{split}
\end{align}
For the mixed components, up to order $\mathcal{O}(v^3)$, using the PN gauge \eqref{PN_gauge}, the PN tensor field equations \eqref{R0j_right} and \eqref{R0j_left} follow 
\begin{align}
\begin{split}
\nabla^2\ac{(3)}h_{0j}+\f{1}{2}\ac{(2)}h_{00,0j}=16\pi G\sum_am_av_a^j\delta^3(\mathbf{r}-\mathbf{r}_a)\,,
\end{split}
\end{align}
and using Eq. \eqref{h_00_2}, the solution is given by
\begin{align}
\begin{split}
\ac{(3)}h_{0j}=-\frac{7}{2}\sum_a\!\frac{Gm_av_a^j}{r_a}-\frac {1}{2}\sum_{a}\!\frac{Gm_a}{r_a^3}(\mathbf{r}_a\!\cdot\! \mathbf{v}_a)(r^j\!-\!r_a^j)\,.
\end{split}
\end{align}
Now, considering the PN tensor field equations \eqref{R00_right} and \eqref{R00_left} up to order $\mathcal{O}(v^4)$, using the PN gauge \eqref{PN_gauge}, and we obtain the equation
\begin{align}
\begin{split}
&\nabla^2\ac{(4)}h_{00}+\f{1}{2}\nabla^2\ac{(2)}h_{00}^2-\ac{(2)}h_{00}\nabla^2\ac{(2)}h_{00}-\ac{(2)}h_{jk}\ac{(2)}h_{00,jk}=
\\&-\!8\pi G\!\sum_a\!m_a\delta^3(\mathbf{r}\!-\!\mathbf{r}_a)\!\Big(\!\f{3}{2}v^2_a\!-\!\accentset{(2)}h_{00}\!-\!\f{1}{2}\accentset{(2)}h_{ij}\delta_{ij}\!+\!\f{\epsilon_a}{2\Mpl}\accentset{(2)}\varphi\Big)\,,
\end{split}
\end{align}
and this solution is derived in the form by applying the above results [\eqref{h_00_2}, \eqref{h_ij_2}, and \eqref{multibody_varphi}],
\begin{align}
\begin{split}
\ac{(4)}h_{00}=&-2\bigg(\sum_{a}\frac{Gm_a}{r_a}\bigg)^2+3\sum_a\frac{Gm_av_a^2}{r_a}
\\&-2\sum_a\sum_{b\ne a}\frac{G^2m_am_b}{r_ar_{ab}}\left(1+\frac{1}{2}\ep_a\ep_be^{-m_sr_{ab}}\right)\,.
\end{split}
\end{align}
In the all above expressions, $m_a$ is the mass of the $a$-th object, $\ep_a$ is its screened parameter, $v_a$ is its velocity, $m_s$ is the effective mass of the scalar field, and $r_a=\left|\mathbf{r}-\mathbf{r}_a(t)\right|$, $r_{ab}=\left|\mathbf{r}_a(t)-\mathbf{r}_b(t)\right|$.
Finally, summing the relevant components of $g_{\mu\nu}$, and we obtain the results presented in Eqs. \eqref{PN_metric}.

\section{Evaluation of integrals arising in the derivation of scalar radiation}\label{appendix_c}
Here, we derive the integrals related to the scalar energy flux.
It is hopeless to get an exact result of these integrals. However, we can obtain
their asymptotic behavior in the wave zone ($r\rightarrow +\infty$).
\begin{eqnarray}\label{2}
  \langle\cos(\omega r u)\rangle_n \!=\!\int_0^\infty \!\cos\left(\omega r \sqrt{1\!+\!(\frac{z}{m_s r})^2}\right)\frac{J_1(z) dz}{(1\!+\!(\frac{z}{m_s r})^2)^{\frac n 2}},
  \nonumber\\
  \langle\sin(\omega r u)\rangle_n \!=\!\int_0^\infty\! \sin\left(\omega r \sqrt{1\!+\!(\frac{z}{m_s r})^2}\right)\frac{J_1(z) dz}{(1\!+\!(\frac{z}{m_s r})^2)^{\frac n 2}}.
  \nonumber\\
\end{eqnarray}

We only discuss the evaluation of $\langle\cos(\omega r u)\rangle_n$, since $\langle\sin(\omega r u)\rangle_n$ can be evaluated in the same way. Choose a parameter $\lambda$ such that
$m_s r \lambda\gg 1$ and split the integral into two parts. The asymptotic expansion of the first part can be obtained by performing integration by parts as follows:
\begin{eqnarray}\label{1}
  &&\int_0^{m_s r \lambda} \cos\left(\omega r \sqrt{1+(\frac{z}{m_s r})^2}\right)\frac{J_1(z) dz}{(1+(\frac{z}{m_s r})^2)^{\frac n 2}}\nonumber\\
  &&= \cos(\omega r)-J_0(m_s r \lambda) \cos(\omega r \sqrt{1+\lambda^2})\!+\!\cdots\,.
\end{eqnarray}
For the second part, when we perform integration by parts, we can exactly cancel the terms in Eq. \eqref{1} that depend on $\lambda$. Therefore, all the contribution that comes from the end point $m_s r\lambda$ can be ignored.

We can substitute the leading asymptotic behavior of Bessel function
\begin{equation}
  J_\nu(x) \sim \sqrt{\frac{2}{\pi x}}\cos(x-\frac{\nu \pi}{2}-\frac{\pi}{4}),
\end{equation}
into the second part, then the integral can be approximated by
\begin{equation}
  \sqrt{\frac 2 \pi}\int_{m_s r \lambda}^\infty \frac{\cos\left(\omega r \sqrt{1+(\frac{z}{m_s r})^2}\right)\cos(z-3\pi/4)}{\sqrt{z}(1+(\frac{z}{m_s r})^2)^{\frac n 2} }dz.
\end{equation}
The above integral can be transformed into complex integral
\begin{equation}
  I=\frac14\sqrt{\frac2 \pi}\int_{m_s r\lambda}^\infty \frac{e^{\rho(z)} dz}{\sqrt{z}(1+(\frac{z}{m_s r})^2)^{\frac n 2}},
\end{equation}
with
\begin{equation}
  \rho(z)= i n_1 \omega r \sqrt{1+(\frac{z}{m_s r})^2}+i n_2(z-\frac34 \pi),
\end{equation}
where $n_{1,2}=\pm 1$. The integration contour that gives the dominant contribution of the integral is determined by $\rho(z)$ and the relative sizes of $\omega$ and $m_s$.

When $\omega>m_s$ and $n_1=-n_2$, $\rho(z)$ has a stationary point at $a=\frac{m_s^2 r}{\sqrt{\omega^2-m_s^2}}$. Using the method of stationary phase \cite{Bender1987}, $I$ can be approximated by
\begin{eqnarray}
  I\sim &&\frac14\sqrt{\frac2\pi}\frac{e^{\rho(a)}}{\sqrt{a}(1+(\frac {a}{m_s r})^2)^{\frac n 2}}\int_{-\infty}^{+\infty} dt e^{\rho''(a)t^2/2}\nonumber\\
  \sim && -\frac12 e^{i n_1\sqrt{\omega^2-m_s^2}r}\left(\frac{\sqrt{\omega^2-m_s^2}}{\omega}\right)^{n-1}.
\end{eqnarray}
When $\omega>m_s$ and $n_1=n_2$, $\rho(z)$ has no stationary point and hence $I$ does not give contribution to the leading asymptotic behavior of Eq. \eqref{2}. All in all, the leading asymptotic behavior of $\langle \cos (\omega r u) \rangle_n$ and $\langle \sin (\omega r u) \rangle_n$   for $\omega>m_s$ are
\begin{eqnarray}
&\langle \cos (\omega r u) \rangle_n \sim \cos(\omega r)\!-\!\left(\!\frac{\sqrt{\omega^2\!-\!m_s^2}}{\omega}\right)^{n\!-\!1}\!\!\!\cos(r\sqrt{\omega^2\!-\!m_s^2}),
  \nonumber\\
&\langle \sin (\omega r u) \rangle_n \sim \sin(\omega r)\!-\!\left(\!\frac{\sqrt{\omega^2\!-\!m_s^2}}{\omega}\right)^{n\!-\!1}\!\!\!\sin(r\sqrt{\omega^2\!-\!m_s^2}).
  \nonumber\\
\end{eqnarray}

When $\omega<m_s$, $\rho(z)$ has two saddle points. Using the method of steepest descent \cite{Bender1987}, we can obtain the leading asymptotic behavior of Eq. \eqref{2}\,,
\begin{eqnarray}
  &\langle \cos (\omega r u) \rangle_n \sim \cos(\omega r)-\left(\frac{\sqrt{m_s^2-\omega^2}}{\omega}\right)^{n-1}
  \nonumber\\&\quad~~~~~~~~~~~~~~~~
 \times e^{-r \sqrt{m_s^2-\omega^2}}\frac{i^{n-1}+(-i)^{n-1}}{2},
  \nonumber\\
 & \langle \sin (\omega r u) \rangle_n \sim \sin(\omega r)-\left(\frac{\sqrt{m_s^2-\omega^2}}{\omega}\right)^{n-1}
  \nonumber\\&\quad~~~~~~~~~~~~~~~~
 \times e^{-r \sqrt{m_s^2-\omega^2}}\frac{i^{n-1}-(-i)^{n-1}}{2}.
\end{eqnarray}

\end{document}